# From 5G RAN Queue Dynamics to Playback: A Performance Analysis for QUIC Video Streaming

Jashanjot Singh Sidhu, *Member, IEEE,* Jorge Ignacio Sandoval, *Member, IEEE,* Abdelhak Bentaleb, *Member, IEEE,* and Sandra Céspedes, *Senior Member, IEEE*

***Abstract*—The rapid adoption of QUIC as a transport protocol has transformed content delivery by reducing latency, enhancing congestion control (CC), and enabling more efficient multiplexing. With the advent of 5G networks, which support ultra-low latency and high bandwidth, streaming high-resolution video at 4K and beyond has become increasingly viable. However, optimizing Quality of Experience (QoE) in mobile networks remains challenging due to the complex interactions among Adaptive Bit Rate (ABR) schemes at the application layer, CC algorithms at the transport layer, and Radio Link Control (RLC) queuing at the link layer in the 5G network. While prior studies have largely examined these components in isolation, this work presents a comprehensive analysis of the impact of modern active queue management (AQM) strategies, such as RED and L4S, on video streaming over diverse QUIC implementations—focusing particularly on their interaction with the RLC buffer in 5G environments and the interplay between CC algorithms and ABR schemes. Our findings demonstrate that the effectiveness of AQM strategies in improving video streaming QoE is intrinsically linked to their dynamic interaction with QUIC implementations, CC algorithms and ABR schemes—highlighting that isolated optimizations are insufficient. This intricate interdependence necessitates holistic, cross-layer adaptive mechanisms capable of real-time coordination between network, transport and application layers, which are crucial for fully leveraging the capabilities of 5G networks to deliver robust, adaptive, and high-quality video streaming.**

## I. Introduction

With the proliferation of online media, ensuring efficient and high-quality content delivery has become a top priority for service providers. Video streaming accounts for a significant portion—around 65%—of global internet traffic, as highlighted in the latest Sandvine report [1]. Given its dominant share, video streaming is not only a critical component of the digital ecosystem but also a key driver of user experience and network performance. Consequently, ensuring seamless delivery of high-quality video is crucial for improving the QoE in an increasingly competitive and bandwidth-intensive environment.

Major industry players, such as Meta, YouTube, and Netflix, rely on over-the-top (OTT) platforms that utilize HTTP Adaptive Streaming (HAS) [2] for content delivery. These platforms dynamically adjust video quality based on network conditions, ensuring optimal playback experiences across diverse devices and varying network environments. For decades, Transmission Control Protocol (TCP) [3] was the preferred transport protocol. However, in recent years, QUIC [4] has emerged as a compelling alternative, replacing TCP as the protocol of choice. Initially developed by Google and later standardized by the IETF, QUIC is a transport-layer network protocol built on top of UDP. With its advanced features, including reduced latency, faster connection establishment, and improved security, QUIC has become the go-to solution for modern, high-performance applications, especially in video streaming and real-time communications [5], [6]. While protocols like QUIC and adaptive streaming technologies have transformed the application layer, advancements at the infrastructure level—particularly in mobile and wireless networks—are equally critical to sustaining high-quality video delivery.

On the infrastructure front, recent advancements in 5G technology have enabled the seamless delivery of 4K (Ultra-HD) video streaming, driving the demand for transferring and processing large volumes of data in real-time. These developments have introduced new requirements for bandwidth-intensive and latency-sensitive applications, which have become essential for the next generation of communication networks. In this context, mobile networks have emerged as a key access point, leveraging their dynamic nature to serve as the universal gateway to the Internet and new applications. As a result, mobile networks have evolved to meet these growing demands, shaping their development around the use cases and services they must support.

The evolution of mobile networks has increasingly relied on high-frequency bands to meet the need for high bandwidth and low latency. This shift has facilitated previously unimaginable speeds. However, a critical question arises: Can these high transfer rates be maintained when users are on the move? What are the characteristics of access networks in such scenarios, and how do they impact applications and services?

The use of millimeter wave (mmWave), coupled with mobility, requires constant adaptation to varying environmental conditions to mitigate interference and signal loss. This adaptation often involves using lower-order modulation schemes and retransmission techniques to safeguard data transmission, resulting in fluctuating bandwidth, delays, and jitter. These variations can lead to network congestion and affect the performance of upper-layer protocols and applications. Consequently, it is crucial to understand how the transport layer can effectively leverage the available network capacity while ensuring that applications continue to perform optimally. To mitigate the congestion effects in the application flow, a buffer is implemented at the RLC layer of the Next-Generation Node

*The first two authors contributed equally to this work.

J. Singh Sidhu, A. Bentaleb and S. Céspedes are with CSSE Department, Concordia University, Montréal, Canada (e-mail: jashanjot.sidhu@mail.concordia.ca, abdelhak.bentaleb@concordia.ca, sandra.cespedes@concordia.ca).

J. Ignacio Sandoval is with University of Chile (e-mail: jsandova@uchile.cl).

B (gNB), as an entry point of the Distributed Unit (DU) entity inside the distributed architecture of the gNB. A large buffer prevents the drops related to short periods of congestion, but when congestion is more permanent, the increased buffering contributes to an increased delay, decreasing the performance of data flows.

CC algorithms and ABR schemes work in tandem to optimize the end-to-end performance of modern streaming networks, with CC algorithm operating at the server side and ABR at the player side. CC algorithm is responsible for managing the flow of data across the network, dynamically adjusting to congestion and preventing overloads, ensuring fair, and efficient use of available bandwidth. On the player side, ABR continuously adjusts the video quality based on real-time network conditions, providing a seamless streaming experience even when bandwidth fluctuates.

However, while CC and ABR are crucial, they are not always sufficient to tackle the full complexity of modern network challenges [7]–[9]. AQM within the network infrastructure, specifically in the gNB RLC buffer [10]–[13], becomes essential. AQM manages packet queues, potentially reducing congestion and delays to ensure a smooth data transmission. When combined with CC and ABR, AQM is expected to ensure an efficient use of resources, reduce latency, and create an overall enhanced user experience.

The primary objective of this work is to evaluate how AQM within the RLC buffers of 5G gNBs influences QUIC-based video streaming performance. We concentrate on key variables including different QUIC implementations—such as QUINN, S2NQUIC, NGTCP2, and MVFST—each designed with distinct architectural features and CC algorithms. Our study also considers the role of Explicit Congestion Notification (ECN) marking [14] and diverse ABR strategies deployed at the player side. By systematically examining these factors, we aim to accurately describe how AQM, when integrated with specific QUIC implementations, CC algorithms, and ABR schemes, creates a synergistic effect that enhances streaming robustness and adaptability. This comprehensive approach provides practical insights for 5G operators to optimize buffer management and improve video delivery quality in highly variable mobile environments. The main contributions of this paper are:

- We present a comprehensive performance evaluation of modern AQM algorithms—specifically CoDel [15] and L4S [16]—compared to the traditional RED algorithm within the 5G RLC buffer. Our study focuses on their impact on QUIC-based video streaming, analyzing how each AQM affects critical metrics such as latency, packet loss, and streaming quality under realistic mobile network conditions.
- We perform a detailed analysis of the interplay among AQM schemes, diverse QUIC implementations [17] (including MVFST, NGTCP2, QUINN, and S2NQUIC), CC algorithms (e.g., BBR [18] and CUBIC [19]), and ABR strategies (such as learning-based Pensieve [20] and throughput-driven methods). This comprehensive examination uncovers complex synergistic and antagonistic interactions that influence overall streaming performance, providing key insights for optimizing cross-layer coordination.
- Through extensive experiments reflecting varying mobility patterns, signal qualities, and traffic loads, we identify optimal AQM parameter configurations that maximize QoE metrics. These include reductions in startup delay, rebuffering occurrences, and improvements in video quality stability, offering practical recommendations for 5G operators to better tailor RLC buffer management to the dynamic characteristics of mobile broadband.
- We quantify the benefits of incorporating ECN marking in conjunction with AQM in the RLC buffer. Our results demonstrate that ECN-enabled AQM significantly enhances congestion feedback to QUIC transport layers, enabling a more responsive CC algorithm that reduces latency and packet loss while facilitating smoother and more stable bitrate adaptation. This integration ultimately leads to a markedly improved user experience in QUIC-based video streaming over mobile networks.

TABLE I: Relevant Acronyms table

| Acronym | Description |
|---|---|
| ABR | Adaptive Bit Rate |
| AQM | Active queue management |
| CC | Congestion control |
| ECN | Explicit Congestion Notification |
| gNB | Next-Generation Node B |
| HAS | HTTP Adaptive Streaming |
| LoS | Line-of-Sight |
| QoE | Quality of Experience |
| MAC | Medium Access Control |
| NLoS | Non-Line-of-Sight |
| RD | Rebuffering duration |
| RLC | Radio Link Control |
| sRTT | Successful round-trip time |
| VBR | Variable Bit Rate |
| VMAF | Video Multimethod Assessment Fusion |

The remainder of this paper is organized as follows. Section II presents the problem description and related works. Section III explains the methodology used in this research. Section IV presents the simulation results and Section V highlights the key takeaways and discussions of our analysis and findings. Finally, we conclude the work in Section VI and identify the next steps. For a complete list of acronyms used in this work, the reader is referred to Table I.

## II. Background and Related Works

### A. Problem Description

The exponential growth of Internet traffic and new throughput-demanding and delay-sensitive applications constrain access networks and transport protocols. Thus, 5G-Advanced and 6G (5G-A/6G) mobile networks play an essential role in providing high bandwidth and low latency. To meet these stringent performance requirements, the systems must rely heavily on utilizing new high-frequency spectrum. The implementation of new technology in the Frequency Range 2 (FR2) (24.25 GHz–52.6 GHz) represents a significant challenge for new networks operating within the traditional sub-6 GHz Frequency Range 1 (FR1) (410 MHz–7.125 GHz). However, utilizing these higher frequencies, particularly FR2 (mmWave), introduces significant propagation challenges compared to traditional FR1 bands, leading to unpredictable performance.

Sudden changes in channel conditions appear as a consequence of mobility and obstruction, leading to highly fluctuating scenarios that affect the performance of upper-layer protocols and services, and threaten the quality promised by the technology. It is possible to have more than 20 dB loss associated with using mmWave according to the channel model used [21]. As shown in Fig. 1, a user walking across two buildings obstructing the signal has a high decrease of the Signal-to-Interference-plus-Noise-Ratio (SINR) and an increase of Block Error Rate (BLER), because the obstruction and this effect is the largest in the highest frequencies. These effects are not normally seen in traditional mobile systems operating in sub-6 GHz bands.

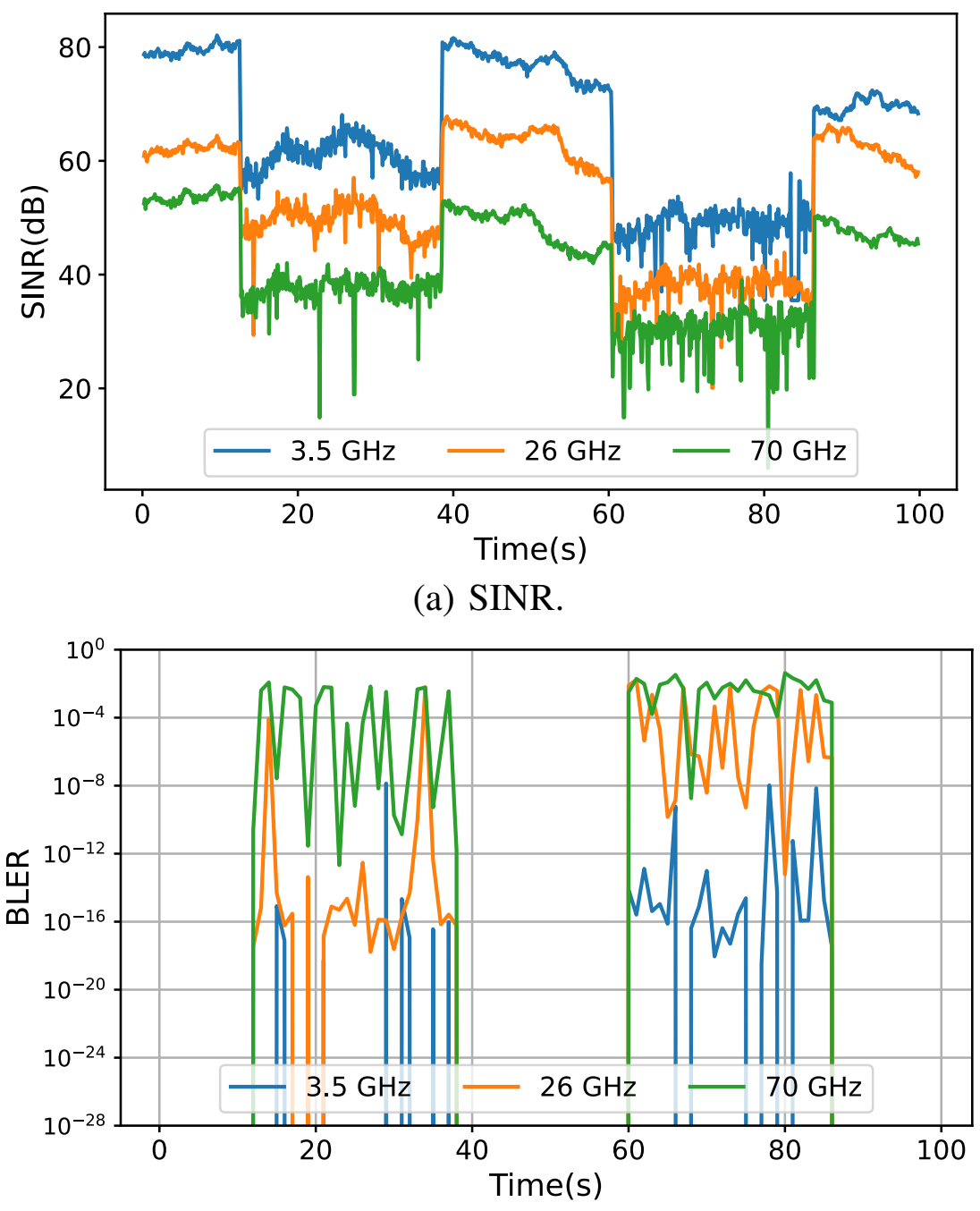

(a) SINR.

(b) BLER.

Fig. 1: Comparison of the impact on the SINR and BLER for transmissions over medium and high-frequency bands. The UE mobility corresponds to a pedestrian walking through a LoS-NLoS environment generated by building obstructions.

When the User Equipment (UE) perceives a decrease in SINR, it switches to a low-order Modulation and Coding Scheme (MCS), decreasing the bandwidth available up to two times and using more spectrum to send the same bits; as a result, there is an increased probability of congestion in the media access. At the same time, an increase in the BLER at the physical (PHY) layer, and retransmissions at the Medium Access Control (MAC) layer, increases the delay as jitter as a product of the random losses in the media access.

In wireless and mobile networks, it is common for congestion to be the result of the fluctuation of the capacity available in the media access or due to new users joining a cell, and not only because of traffic bursts, as in fixed networks. It is important to note that users who are at the limits of coverage or whose signals are obstructed by obstacles transmit information using lower modulation orders, so they require more time on the channel than those who are closer to the cell and can use higher modulation orders, which creates an environment conducive to congestion. [22].

### B. Related Work

*1) Endpoint Congestion Control Algorithms:* The transport layer's CC algorithm are the traditional way to solve the congestion problem in packet-switched networks. CC algorithms normally react to congestion through adaptation of the data rate. To prevent packet losses due to bursty traffic, network devices implement buffers; nevertheless, when the buffers are full, nodes proceeds to drop packets.

The traditional additive increase and multiplicative decrease policies [23], in use for more than four decades, have the objective of keeping the flow at the operational point near the capacity network, reaching a steady state. When the operational point changes due to a decrease in the offered load or an increase in the available capacity, there is a need for mechanisms that move the flow capacity to the new operational point. Loss-based algorithms use packet losses as a congestion signal; this means, losses are interpreted as the result of flows using the available capacity and filling up buffers at network devices, incrementing the delay and eventually triggering packet losses. In most cases, with the congestion avoidance mode, the algorithm appropriately reacts to such losses. However, other congestion detection mechanisms, such as the *probe phase* introduced in the bottleneck bandwidth and round-trip propagation time (BBR) algorithm [24], are based on the monitoring of changes and the adaptation of throughput to new network states, optimizing as a result the throughput and delay of traffic flows.

The emergence of QUIC [4] has prompted various efforts to develop tailored implementations for specific performance and scalability needs. These implementations tackle issues such as the head-of-line blocking problem and enable flow multiplexing, thereby enhancing performance—particularly when transmitting small volumes of data across multiple streams.

LSQUIC [17], by LiteSpeed, focuses on low latency for web browsing and streaming, while TQUIC [17], from Tencent, targets low latency in gaming and live services. mvfst [17], by Meta, emphasizes reliability for large-scale social media deployments, and QUICHE [17], from Cloudflare, prioritizes secure, high-performance web services. QUINN [17], implemented in Rust, excels in managing multiple concurrent connections efficiently. These implementations vary in the supported CC algorithms: mvfst offers BBR and CUBIC [25], QUICHE includes Reno and CUBIC, and LSQUIC provides flexible options, enhancing the QUIC ecosystem's adaptability to diverse applications, from real-time communication to high-quality media streaming.

*2) Network Congestion Control Algorithms:* The first technique to prevent congestion and packet losses from burst traffic is the implementation of buffers in the network. When the data rate of the flow exceeds the bottleneck bandwidth, the buffers fill up and may start dropping packets. Active queue management (AQM) is a CC mechanism designed and implemented in network devices to reduce physical congestion and improve end-to-end latency, complementing endpoint congestion avoidance mechanisms. The traditional technique for managing the queue in a network device is to set a maximum length (in terms of packets) for each queue. Until this maximum length is reached, packets are accepted into the queue, and it is reduced by the transmission of a packet. Subsequent incoming packets are rejected (dropped) until the queue length is within the limits. This technique is known as drop-tail. The packet dropped is detected when the retransmission time-out (RTO) happens and is interpreted by the server as a congestion sign, and the algorithm adjusts the data rate to the new condition. In this sense, some mechanisms have been developed to improve the user flows, detecting the congestion early and dropping packets or sending ECN to the server.

The first generation of AQM mechanisms worked mainly on the queuing process, with strategies to discard incoming packets, such as the RED family. The next generation of AQM attempted to analyze the waiting time of each packet in the queue and discard the packets in the output process, such as the CoDel family.

Random Early Detection (RED) is a congestion avoidance queue management algorithm developed by Floyd and Jacobson in 1993 [26]. RED monitors the average queue size and drops packets, based on statistical probabilities, before the queue is completely full. The mechanism calculates the average queue size ($avg$) using a low-pass filter with an exponential weighted moving average. It is compared to a minimum $min_{th}$ and a maximum $max_{th}$ threshold. When the $avg$ is less than $min_{th}$, packets are enqueued. Any arriving packet is dropped when the $avg$ is greater than $max_{th}$. Finally, if $avg$ is between $min_{th}$ and $max_{th}$, every incoming packet is dropped with a probability $p_a$, where $p_a$ is a function of $avg$.

The drop probability is calculated according to (1) and (2):

$$p_b = p_{max} \times (avg - min_{th})/(max_{th} - min_{th}), \quad (1)$$

$$p_a = p_n/(1 - count \times p_b), \quad (2)$$

where $p_{max}$ is the maximum probability of drop packets, and $count$ is the number of packets enqueued since the last drop.

The congestion state starts when the $avg$ usage is over the threshold $min_{th}$ and the packets are dropped with a probability $p_a$, and the sensibility to traffic burst is defined by the calculus of the $avg$. In this case, the packets are dropped before being enqueued in the buffer. The $avg$ is calculated using the exponential weighted moving average method according to (3), when $oldavg$ is the previous $avg$ and $current_queue_len$ is the current length of the queue. This method allows us to define when the mechanism starts to work; assigning an appropriated queue weight ($w_q$) makes it possible to define the contribution of the previous average ($oldavg$).

$$avg = (1 - w_q) \times oldavg + w_q \times current_queue_len \quad (3)$$

In [11], Anup *et al.* analyze the effects of AQM-based RLC buffer management on the performance of different MAC schedulers. In this work, the authors show that the application of Random Early Detection (RED) may improve the throughput in LTE networks. Nevertheless, the study uses LTE in the FR1 band (from 450 MHz to 6 GHz); therefore, the response to movement and obstructions is not analyzed for mmWave frequencies at 5G-NR.

In [27], Feng *et al.* showed that the effectiveness of RED depends mainly on the proper parametrization of the queue, and proposed a self-configuring RED [27]. Floyd *et al.* further elaborates on Feng's work and proposes Adaptive RED (ARED) [28], which adjusts the maximum dropping probability, $max_p$, to keep the $avg$ between $min_{th}$ and $max_{th}$. It also makes several significant algorithmic changes to the original Adaptive RED proposal. ARED, by dynamically adjusting the maximum drop probability ($max_p$), and automatically tuning the queue weight ($w_q$) and the maximum threshold ($max_{th}$), ARED maintains a predictable average queue size while mitigating the sensitivity of traditional RED. In the same way as RED, the congestion state starts when the $avg$ usage is over the threshold $min_{th}$ and the packets are dropped with a probability $p_a$, and the sensitivity to traffic burst is defined by the calculus of the $avg$. The big difference is that in this case, the maximum probability of dropping is adaptive, then the impact can be more significant.

In [9], [13], we analyze the state-of-the-art in CC algorithm and AQM, such as RED and ARED, and introduce a new performance metric—the convergence time—to analyze the performance of CC algorithm over fluctuating conditions in mmWave links. However, the impact of the interplay between adaptive application data rates, reactive actions from transport-layer CC, and buffer management at the RLC remained unexplored.

Packet networks require buffers to absorb short-term bursts, but loss-based CC algorithm generate a persistently full buffer problem, increasing the delay. Nichols and Jacobson proposed Controlled Delay Management (CoDel) as a solution to avoid buffer bloat [15], [29]. CoDel is an algorithm that works by discarding packets if the time in the buffer is persistently too long ($qdelay_{target}$). CoDel is the first AQM algorithm to use queue delay as a congestion metric, as described in RFC 8289 [15]. CoDel works during the dequeuing process when a packet is exiting through the output port. In the same line, the Flow Queue CoDel (FQ-CoDel) algorithm [30] combines the usage of multiple queues and a Deficit Round Robin Scheduler with CoDel on each queue.

CoDel defines a congestion state when the packets dequeued from the buffer have a sojourn delay over a $target$ for more than an $interval$ of time. At this moment, all packets are dropped in the dequeue process. In [12], Irazabal *et al.* propose a bufferbloat avoidance algorithm and a scheduler for circumventing the added sojourn time caused by the packet segmentation/reassembly procedure of the RLC layer. The authors used a testbed with CUBIC as the CC, to analyze the impact of the added delay due to segmentation and reassembly as a way to fill Resource Block gaps and optimize link utilization. This work opens the questions to analyze the impact of the RLC buffer and the usage of AQM as CoDel, to manage the sojourn time.

*3) Explicit Congestion Notification (ECN):* The ECN [31] is a mechanism that indicates congestion in the network. It is used by the AQM mechanisms to provide an indication of this congestion to the end nodes. The AQM use the Congestion Encountered (CE) codepoint in the IP packet header as an indication of congestion. This mark is received by the client and replied to the server in the next acknowledgment (ACK) and can be interpreted as congestion and update the data rate of the flow. In the case of TCP, it is necessary first to set up an ECN capable session in the synchronization phase, and then it is possible to replay with a ECN-Echo (ECE) flag of the reserved bytes of TCP.

The usage of this mechanism at the gNB requires working in different layers. According to the mobile architecture, a buffer is implemented in the RLC layer, where the packets are dequeued and sent through the MAC and PHY layers to the users. However, the packets in the RLC layer are encrypted; thus, marking has to be placed in the previous Packet Data Convergence Protocol (PDCP) layer.

Under normal conditions, losses are interpreted as a sign of congestion and a multiplicative decrease in congestion window (cwnd) is applied, but if the algorithm is compatible with ECN, we will have a new congestion signal and if it receives an ECE flag, this can be interpreted as a soft signal of congestion and *cwnd* can be reduced with less effect than a packet loss.

In [32], Alwahab and Laki analyze the usage of the ECN with CoDel and show how to reduce the number of retransmissions in classic CC algorithm such as CUBIC and Reno. The authors chose a dumbbell topology, designed and built in Mininet, and analyzed the performance of different variants of TCP, such as Reno, CUBIC, and BBR, over this scenario and the usage of CoDel as an AQM mechanism, and then compared the effect of ECN. Despite the simple architecture, it was possible to see the improvement in the usage of the mechanism of marking over different congestion stages based on changes in the last-mile link delay. The research is based on a fixed network and does not include the fluctuating effects resulting from mobility, obstructions, and high throughput demands.

In [33], authors analyze the usage of ECN in BBRv2 and show the improvement in fairness without sacrificing the algorithm's performance. The AQM implemented works analyzing the queue length and marking packets with a probability that changes linearly with the queue length and the buffer usage between a defined threshold. The scenario chosen is a dumbbell topology simulated over ns-3.

Low Latency Low Loss Scalable (L4S) proposed in [10] is an algorithm that works marking packets using ECN base on the queue delay ($q_{delay}$) of each packet. When the $q_{delay}$ is less than $l4sLowTh$, packets are dequeued. Any enqueued packet is marked when the $q_{delay}$ is greater than $l4sHighTh$. Finally, if $q_{delay}$ is between $l4sLowTh$ and $l4sHighTh$, every dequeue packet is marked with a probability $pMark$, where $pMark$ is a function of $q_{delay}$.

The marking probability is calculated by the two-step formula:

$$temp(t) = \frac{q_{delay} - l4sLowTh}{l4sHighTh - l4sLowTh} \quad (4)$$

$$pMark = \alpha \times temp(t) + (1 - \alpha) \times (t - dT) \quad (5)$$

Where $temp$ represents the instantaneous marking probability and $\alpha$ is the weights associated with the contribution of the previous probability in the period $dT$.

The congestion state is defined when packets are more than $l4sLowTh$, and the algorithms start to mark them with a probability of $pMark$; this means all next packets in the buffer can be marked up to the $qtime$ decrease below $l4sLowTh$.

In [10], authors implement L4S marking strategy in the PDCP layer of the gNB and evaluate the experimental standard SCReAM [34] for the CC to support Augmented Reality (AR) video gaming traffic. The work shows that it is possible to achieve low latency and lower packet losses while maintaining a good throughput performance. However, it does not measure the impact of the decrease in throughput on the application. This research has a similar approach to ours, as it implements a congestion detector in the RLC layer, where the packets are enqueued, but in this layer, packets are encrypted, then the packet marking is made at the PDCP layer. The testbed used is based on a simulation of a full 5G-New Radio (5G-NR) Network, working at 600 MHz with users distributed randomly in the scenario, with a fixed position in the whole simulation. The users are video gaming users who share access to media with web users in a proportion of 1 to 10.

## III. Methodology

Our methodology employs a hybrid simulation-emulation framework to comprehensively evaluate the impact of RLC-layer AQM strategies on QUIC-based video streaming QoE in dynamic 5G mmWave scenarios. We analyze the interactions between AQM mechanisms, diverse QUIC server implementations, CC algorithms, and ABR schemes under realistic channel conditions. Our architecture, depicted in Fig. 2, uses ns-3/5G-LENA to generate realistic channel traces offline, where a group of users is moving around a gNB, and the signal is obstructed by buildings. Changes in the available capacity, delay, jitter, packet drops, and other significant parameters and metrics are stored and used to characterize the environment and affect traffic between the UE and server with a video streaming application. Then, Vegvisir [35] architecture evaluates the behavior of QUIC-based video streaming over different mechanisms of AQM to improve the performance of the application over a congested channel.

### A. Scenarios of Study

The requirements of the new services point to bandwidth-demanding and delay-sensitive applications, and to satisfy these requirements, new technologies have had to start using higher bandwidths only available at higher frequencies. The usage of mmWave, the user mobility, and the environment obstructions offer a new complex scenario where sudden changes in the capacity offered, delays, and random drops in the media access are challenging for evaluating the interplay of AQM, QUIC, CC, and ABR. Following the scenario proposed in [9], a single gNB was positioned at a height of 15 m and operated at mmWave frequencies. The users walk at a speed of 1 m/s, encountering two buildings that block the communication path between the UEs and gNB (Fig. 3), creating a sequence of Line-of-Sight (LoS) and Non-Line-of-Sight (NLoS) zones. A semi-circular path was chosen, keeping the distance to the gNB constant and thus avoiding the effect of signal attenuation due to distance. In this way, it is possible to connect the users to a server that will provide a data flow and the response of the CC algorithm to the fluctuating scenario can be studied with a set of AQM strategies, such as RED [26], ARED [28], CoDel [15], and L4S [10].

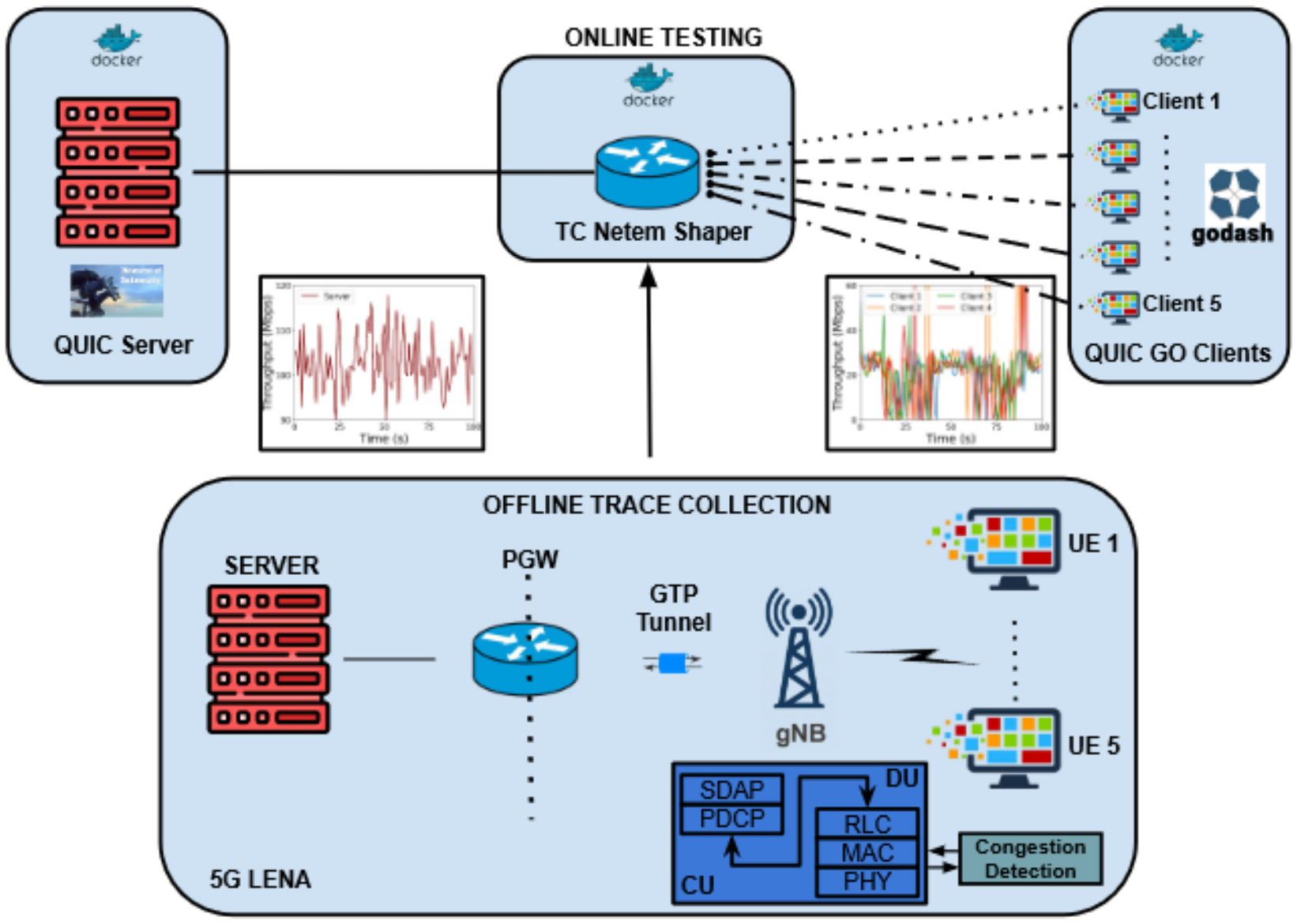

Fig. 2: Experimental setup.

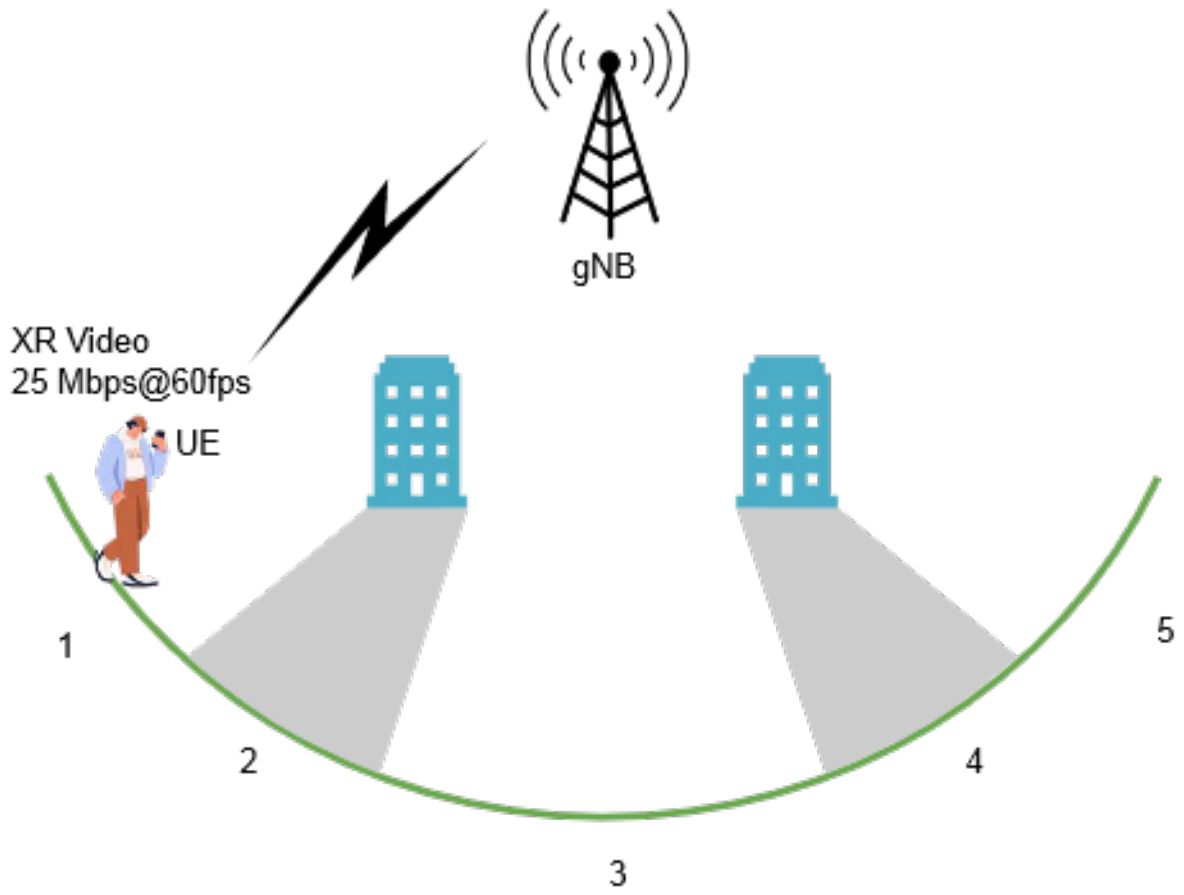

Fig. 3: Simulation scenario for CC evaluation: An outdoor environment with a gNB operating in a mmWave band and UEs moving around it.

## B. Simulation Characteristics

*1) Offline architecture:* The ns-3 framework and the 5G-LENA module [36] were used. 5G-LENA incorporates essential PHY-MAC features from 5G-NR, following Release 15 TS 38.300, and the flexibility to add new models and algorithms in different layers, as well as the inclusion of user mobility and obstacles, such as buildings. 5G-LENA supports a flexible frequency range from 400 MHz to 100 GHz, allowing testing of the NR operation, which was extended to 71 GHz included in Release 17. It supports a loss propagation model that replicates path loss, shadowing, and fading per 3GPP TR 38.901 [21]. Mobility effects, band selection, obstructions, and network usage are exported to an online testbed with real servers, a video player, and an intermediary router as shown in Fig. 2, with an overview of the simulation parameters inside the ns-3 environment presented in Table II. To calculate the RLC buffer size, we used a nominal bandwidth of 500 Mbps and the server delay of 4 ms, and this size is fixed in all the path, including the NLoS area, where the maximum capacity decreases, according to the best equilibrium between maximum throughput and minimum delay, as described in [9].

TABLE II: Offline Simulation Parameters

| Parameter | Value |
|---|---|
| Number of UE | 5@VBR |
| Traffic Model | VBR: XR, 60 fps@25 Mb/s |
| Simulation Time | 100 s |
| Bandwidth & Frequency | 100 MHz @ 27.3 GHz |
| RLC Buffer | 200%BDP |
| Delay to server (one-way) | 4 ms |
| UE speed | 1 m/s |
| RED $min_{th}$ & $max_{th}$ | 90% to 95% & 100% Buffer Size |
| ARED $min_{th}$ | 90% to 95% Buffer Size |
| CoDel $qdelay_{target}$ | 1 to 100 ms |
| L4S $min_{th}$ & $max_{th}$ | 1 to 100 ms |

The ns-3 is a framework developed in C++, and to improve the flexibility, the implementation of AQM mechanisms was made using Python code, and the integration was made with a memory message interface based on [37] that in our implementation we connect the C++ RLC layer with the queue manager in Python.

The gNB has one RLC buffer per flow, and each one implements the management mechanism and works independently of the other. Then, at the MAC layer, a round-robin scheme is used in the scheduler to deliver the packets from each queue to the PHY layer. This means that algorithms such as CoDel behave as FQ-CoDel [30]. In the same way, the use of multiple queues allows for the prevention of the lockout problem [38] when long-lasting flows monopolize the queue.

TABLE III: Traffic Model Parameters

| Parameter | Frame Size | Jitter |
|---|---|---|
| distribution | Truncated Gaussian | Truncated Gaussian |
| mean | $datarate/fps/8$ | $Zero$ |
| std | $0.15 \times mean$ | $2\ ms$ |
| min | 67 Bytes | $-4\ ms$ |
| max | $1.5 \times mean$ | $4\ ms$ |

TABLE IV: QUIC implementation details.

| QUIC Implementation | Language | Multi-Client Handling | Pacing |
|---|---|---|---|
| QUINN | Rust | Asynchronous threads | ✓ |
| LSQUIC | | Multiple threads | ✓ |
| TQUIC | | Multiple threads | ✗ |
| S2NQUIC | | Asynchronous threads | ✓ |
| NEQO | | ✗ | ✓ |
| LSQUIC | | ✗ | ✓ |
| PICOQUIC | C | Multiple threads | ✓ |
| NGTCP2 | | Multiple threads | ✓ |
| MVFST | C++ | Asynchronous threads | ✓ |

The congestion detection function receives status information from the RLC layer, such as the buffer usage and the sojourn delay in the buffer. Based on this information and the AQM mechanism, it is possible to discard packets in the enqueue or dequeue process, according to the mechanism chosen, or mark the congestion in the incoming packets in the PDCP layer before the packet is encrypted. Our implementation allows us to receive multiple inputs to decide the congestion status and take action, such as Channel Quality Indicator (CQI) or the information from the queue of the other users in the same radio base.

The AQM mechanisms used were developed to work dropping packets proactively in the case of RED, ARED, and CoDel and marking packets in the case of L4S. In the experiments conducted, the variants of packet discarding and packet marking were implemented for each algorithm to establish clearer points of comparison.

We tested each AQM mechanism using different parameter settings according to Table II and then selected the one with the best performance for each family.

The traffic model is a Variable Bit Rate (VBR) video traffic; the implementation was made according to the 3GPP recommendation for eXtended Reality (XR) services in [39]. In this document, a truncated Gaussian distribution is recommended to model the frame size of a video service; each frame is sent according to the generation rate, 60 fps, and is added a jitter that can be modeled as another truncated Gaussian distribution, as shown in Table III.

*2) Online Testing:* To evaluate the performance of our setup in an online testing environment, we extended the Vegvisir architecture [35] to support five QUIC clients connected to a single QUIC server. The server used a single interface, while each client was assigned a separate network interface, as shown in Fig. 2. The client-server communication was routed through a centralized traffic shaper (TC Netem), which acted as the link bottleneck. To emulate realistic network conditions, we used trace data collected offline from 5G-LENA, directly feeding this trace into TC Netem to simulate trace-driven 5G conditions. Each component—the QUIC server, clients, and shaper—was encapsulated in its own Docker container. The clients were each linked to the server through their own dedicated Docker network, ensuring isolated, secure connections. This setup created a cohesive, scalable streaming system. To simulate a real-world video streaming session, we emulated an HAS session with 4K video under various network conditions.

TABLE V: Online emulation parameters

| Scen. | ECN Marking | CC | Player Max. Buffer(s) | Seg. Duration (s) |
|---|---|---|---|---|
| A | ✗ | BBR | 6 | 2 |
| B | | CUBIC | | |
| C | ✓ | BBR | | |
| D | | CUBIC | | |

1) Video Contents: We used the Moment of Intensity (MOI) [40] video, encoded using *AVC* codec [41] at 60fps across 10 bitrate levels, ranging from *512x288* to *3840x2160* (4K) with bitrates from 0.5 Mbps to 49 Mbps. The perceptual quality, evaluated through Video Multimethod Assessment Fusion (VMAF) [42], spanned a range of 31 to 97, capturing varying levels of visual fidelity.
2) QUIC Implementations: The client-side utilizes the QUIC-go implementation [17], with the goDASH player running in headless mode to facilitate the extraction of performance metrics. On the server side, given that our work examines the impact of ECN signaling versus no signaling, we selected QUIC implementations that provide full support for ECN. According to the QUIC interoperability tests [43], the following implementations support ECN signaling: NGTCP2, PICOQUIC, NEQO, LSQUIC, QUINN, and S2NQUIC. The details of these QUIC implementations are shown in Table IV. Among these, NEQO and LSQUIC either lack support for handling multiple client connections concurrently or are not fully optimized for such scenarios. During our simulations, we observed that PICOQUIC and TQUIC were unable to complete the simulation events due to excessive rebuffering. This can be attributed to several factors: in PICOQUIC, CUBIC requires multiple stabilization steps following HyStart [44], along with variability in BBR throughput [45], and in TQUIC, the lack of pacing support [46] contributes to the issue. This limits our focus to 4 QUIC implementations on the server-side–MVFST, NGTCP2, QUINN, and S2NQUIC.
3) Congestion Control Algorithms: The 5G LENA model in ns3 supports the following CC algorithms for QUIC with ECN marking: BBR, CUBIC, Reno, and DCTCP. However, since all of these operate over raw QUIC, and based on our online testing, only BBR, CUBIC, and Reno are supported in the QUIC implementations during the testing phase. DCTCP is excluded from our analysis, as it is not supported in the online phase, which could lead to discrepancies in the final results. Among CUBIC and Reno, both of which are loss-based algorithms, we chose CUBIC due to its wider adoption in Linux kernel implementations.
4) Adaptive Bitrate Schemes: We implemented two widely-used ABR algorithms: the heuristic-based conventional

(CON) approach from [47] and the learning-based Pensieve [20]. CON is a throughput-driven ABR method that uses the last two throughput estimates to make bitrate decisions. In contrast, Pensieve dynamically selects the optimal streaming bitrate in real time by considering a range of factors, including network conditions (e.g., bandwidth and latency), video quality metrics, and playback history. To tailor Pensieve to our video's 10 bitrate levels, we retrained the model with additional epochs. Furthermore, we enhanced both Pensieve and CON by incorporating a stall prevention mechanism from [47], resulting in two improved variants: Pensieve$^+$ and CON$^+$.

5) Simulation Setup: The details of the online emulation are presented in Table V. Notably, in Scenarios C and D, where ECN marking is enabled, ECN was active during both the offline and online phases at the QUIC server and the router. Each simulation ran for 100 seconds per AQM strategy and QUIC implementation. Although we tested longer durations, 100 seconds was sufficient to capture the impact of AQM on the RLC buffer and, consequently, user QoE.

### C. Metrics

*1) Throughput:* The throughput measures the link utilization or flow rate in bits per second. When measured in the sender, all packets sent are counted and include the packet product of retransmission. A channel with losses can increase the transmitted throughput product of the retransmissions, but the received throughput can be the same as others without losses. The received throughput or goodput may be used because it only accounts for useful traffic while excluding duplicate packets and those likely to be dropped. Goodput may be a more critical metric, as individual applications often have application-specific throughput requirements beyond mere capacity utilization.

*2) Successful RTT:* In [9], the authors show the impact of retransmission on the delay when it is included in the measurement; therefore, we will call the successful round-trip time (sRTT) as the round trip time it takes for packets to be transferred, including the delay caused by re-transmitting lost packets, i.e. the time between the packet is sent and it is successfully received the first ACK.

*3) Jitter:* The Jitter is the variation in time that takes a packet to arrive across a network. The periodicity of a sent packet is modified by the burst in the network, but in the case of wireless network losses in the access media increment the jitter product of the retransmission in the physical layer.

*4) RLC buffer usage:* It is the level of occupation of each buffer. The usage of buffers allows for the prevention of packet-dropping products from traffic bursts. The size of the buffer defines an increment in the throughput of the data flow but, at the same time, an increment in the delay. Algorithms such as RED use buffer usage to define whether a packet will be dropped.

*5) Queue Delay:* Queue delay (qdelay) is the sojourn time in the queue of the RLC buffer. Algorithms such as CoDel or L4S work based on the qdelay to drop or mark the packets.

*6) Video Multimethod Assessment Fusion (VMAF):* [42] is a perceptual video quality metric developed by Netflix that combines multiple quality features using machine learning. It provides a more accurate representation of human visual perception compared to traditional metrics like PSNR and SSIM, making it widely used for evaluating video streaming quality.

*7) Rebuffering duration (RD):* It is the measure of playback interruptions, and this metric measures how long in seconds a video viewer waits for a video to start to play again.

## IV. Performance Evaluation

### A. Initial Results

The scenario under study was conducted to show the fluctuating bandwidth caused by obstruction and motion. Under such circumstances, the signal experiences fading and shadowing, and it is needed using low order of MCS, this means fewer bits can be sent and an increased BLER, as shown in Fig 1 (b), that contributed to an increase in the losses in PHY layer and this effect as more delay product of the HARQ scheme.

Losses in PHY layer generate an increment of retransmissions due to applying the HARQ scheme. These losses as part of HARQ occurred in media access, and if the maximum number of retries is reached, the algorithm stops to retry and drops the packet, and the upper layers experience a loss. Meanwhile, the upper layers perceive a variable delay because packets spend more time in the queue waiting to be effectively transmitted. Figure 4 shows the effects; although the mean values do not change, it can be seen how the distribution curve is affected and how the effect increases with frequency despite having a channel with available capacity.

These effects added to the changes to low order MCS initiate a period of congestion that can be depicted in Fig. 5.

Fig. 5(a) shows that in the LoS zone, the application flow is stable with increasing cwnd products of the continuous packet acknowledgment. When the UEs enter the NLoS zone at $t_0$, the capacity offered to these users decreases, and the packets start to enqueue in the RLC buffer. In the drop-tail case, when the queue is full, the packets are dropped in $t_1$, and in $t_2$ the sender does not receive the confirmation and starts to not update the cwnd according to the multiplicative criteria or the algorithm definition. The RTO timer is reached in $t_3$, then the congestion state is activated, the lost packets start to be retransmitted, and the cwnd decreases for each packet dropped; the new flow rate is perceived in $t_4$. The buffer will stop dropping the packet when the flow decreases as a result of a lower flow at $t_4$. The flow starts to adapt to the new bandwidth, and the cycle starts again. From the figure, it is possible to observe the number of packets dropped from $t_1$ to $t_5$, and the response of the algorithm to the congestion (effective in $t_4$) depends on the delay to the server. On the other hand, while the algorithm is adapting to a new data rate, all the exceeded traffic will be dropped, and the numbers depend on the new capacity.

Fig. 5(b) shows in $t_1$ and $t_2$ proactive drops as the response to the AQM mechanism. The effect of this drop is a decrease in the cwnd after the RTO timer is reached in $t_3$ and $t_4$, and the effect is perceived when the new flow arrives at the buffer.

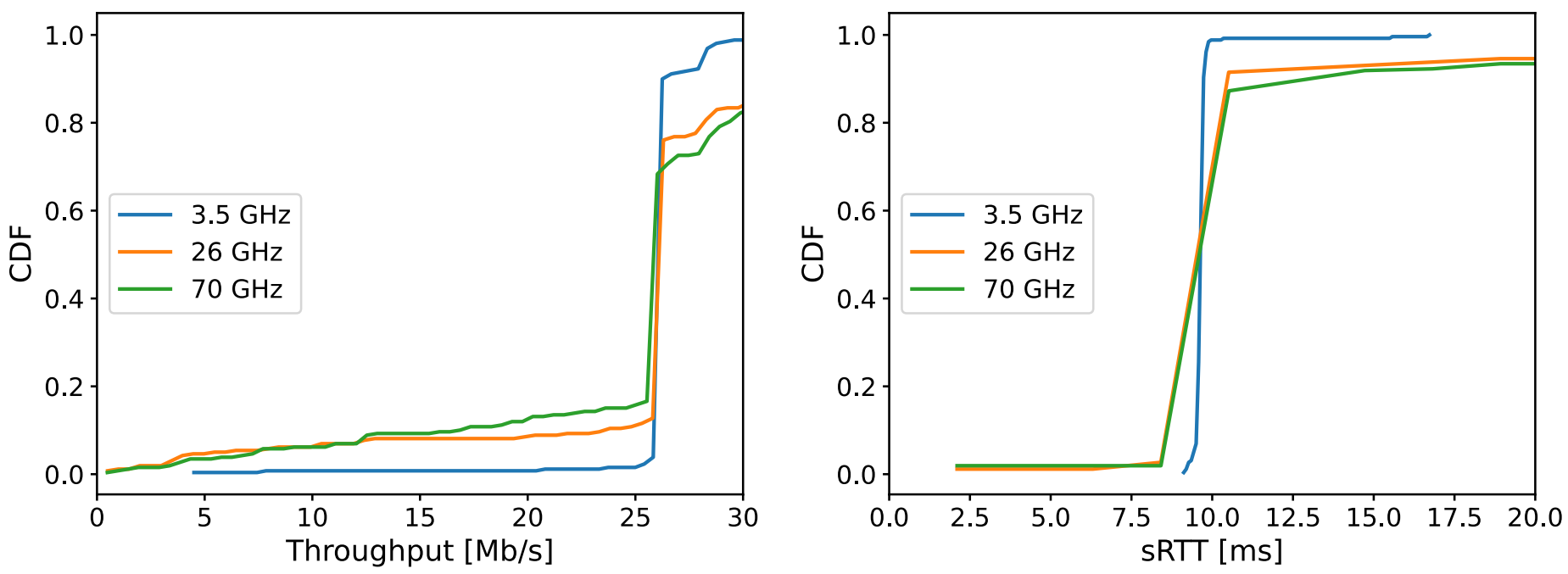

(a) Throughput. (b) Successful RTT.

Fig. 4: Comparison of the effect of high-frequency usage on the throughput and sRTT.

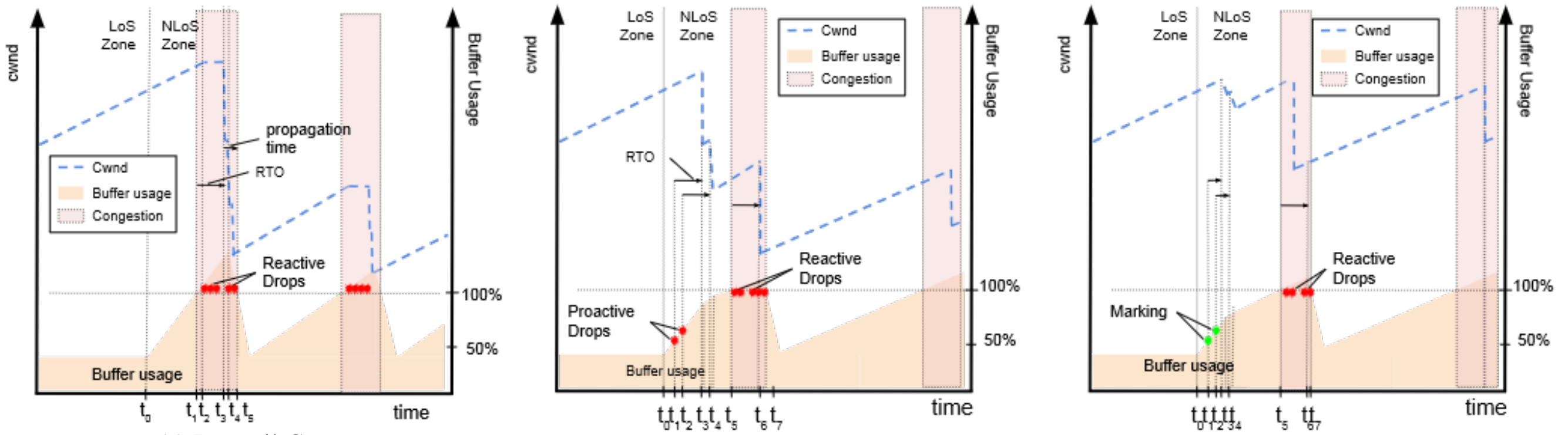

(a) Droptail Case. (b) AQM Case. (c) ECN Case.

Fig. 5: Effects of losses over application flow.

In $t_5$, packets start to be dropped, and after RTO time, the flow is adapted again and perceived in $t_6$, finishing the congestion in $t_7$ when the new flow no longer causes packet dropping.

The proactive drops generated by AQM mechanisms decrease de cwnd early and slow down the flow of packets before filling out the buffer. The immediate effect in the network is fewer packets dropped, but the efficiency depends on when and how many packets are dropped. If the algorithm is too aggressive, the flow could decrease more than expected.

In the case of ECN usage, Fig. 5(c), the congestion signal is replaced by the packet marking with the CE codepoint in $t_1$ and $t_2$. When the mark is received by the server, in $t_3$ and $t_4$, the CC algorithm uses a more conservative response as an additive decrease, and the flow rate decreases for each congestion signal. The buffer fill rate decreases, and the congestion zone is delayed to $t_5$.

All these phenomena are part of a bandwidth fluctuations scenario and are logged and used in the online testing to evaluate the effects on the applications.

### B. *Marking vs Dropping*

The CC algorithm detects the congestion based on a loss detection, and the policy is a multiplicative decrease according to the basic principles described above, and the first reaction occurs after the RTO has elapsed. This means, that when we have an obstruction, the capacity of the access media decreases, and the packets start to be enqueued if the buffer is full, packets start to be dropped, and if the flow of incoming packets is more significant than the sent packet, a train of packet can be dropped before the sender decrease the data rate. In this case, the server will detect all the losses, and the effect in the algorithm can be a large reduction of the cwnd and the data flow.

The AQMs use a proactive drop principle to reduce the flow, but under the same multiplicative decrease principle, the flow will be reduced, but the flow is not interrupted because the next packet is received and the flow is increased according to the next acknowledge sequence optimizing the performance of the connection and the network as was depicted in [13].

The use of ECN allows an improved solution where the usage of the ECE flag to mark a congestion detection is interpreted as a soft drop, and the action in the congestion state machine triggers an action less aggressive than a drop; in the case of DC-TCP, this action is an additive decrease. This means the algorithm decreases the flow throughput without dropping the packets.

Fig. 6 shows different mechanisms used with a DC-TCP flow after entering into the NLoS zone. The star shows the point of the mean throughput and sRTT of the flow of VBR streaming service at 25Mbps and how is reduced as a result of the building obstruction. The x's show the improvement of each AQM mechanism. It is possible to increase the throughput and reduce the sRTT using mechanisms such as RED, ARED, CoDel, or L4S. A second improvement is possible when ECN is used. Arrows show how the operating point is moved. Traditional AQM are highly sensitive to their parameter settings, then a selection of the best AQM performed is shown. RED (illustrated in red), with a min threshold of 80 %, shows the best performance in terms of throughput, although there is a small increment in the sRTT after the use of ECN marking. In the case of ARED (illustrated in orange), a threshold of 90 % is used, it shows a 20% decrease in terms of RTT. In the case of CoDel (illustrated in purple), a threshold of 10 ms is applied, showing the best performance in terms of RTT. This metric decreases more than 28 % with a small reduction of the throughput. In the case of L4S (illustrated in blue), a min threshold of 10 ms and a max threshold of 25 ms is used, and an improvement in both metrics.

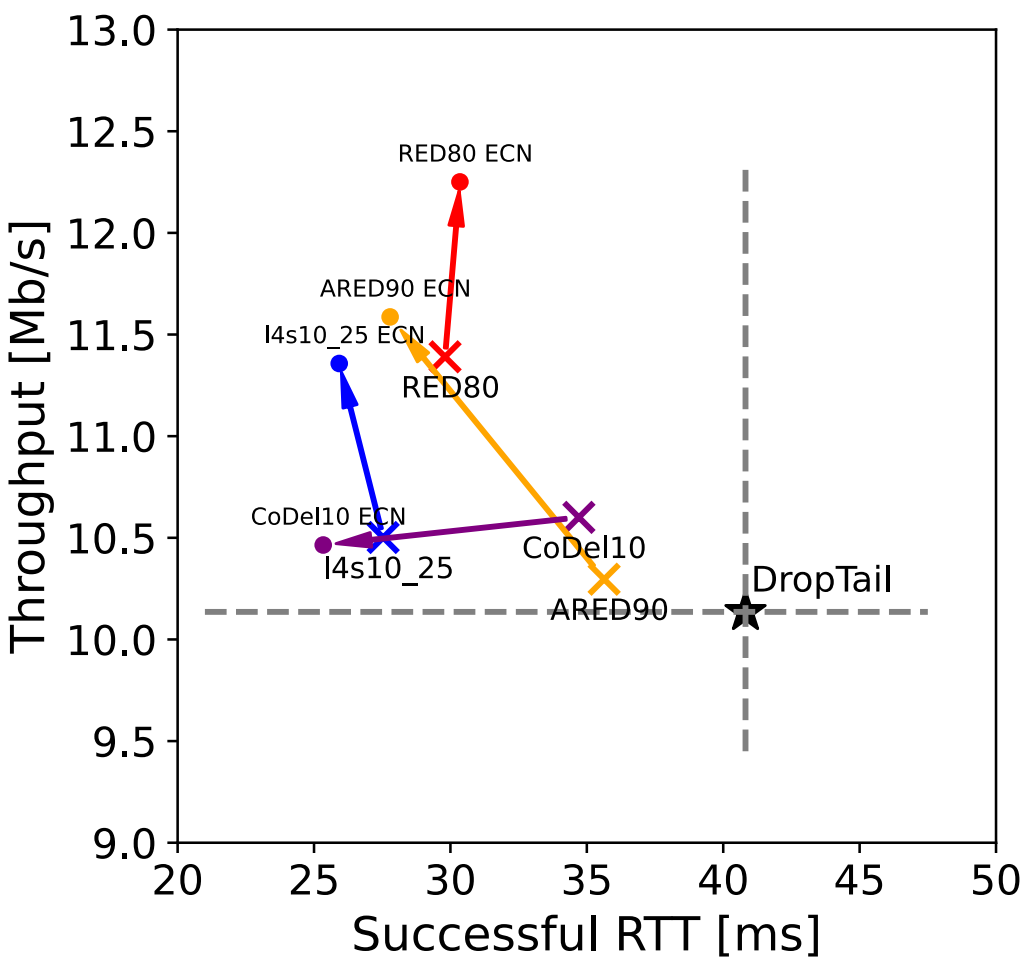

Fig. 6: Comparison of a selection of AQM under DCTCP. The start shows the usage of drop-tail as a baseline. The X shows the throughput and the sRTT of each mechanism. The circles show when ECN marking is used.

Although not all cases were improved after using ECN, we can see that the usage of AQM techniques alone or combined with ECN marking improves the flow performance and increases the efficiency of the gNb. The proactive technique of dropping packets allows for slowing down the throughput by generating more moderate congestion when the buffer is full, in the same way as the packet marking technique by preventing the loss of a train of packets. Now, the proactive technique of dropping packets or marking could have an effect on the applications. These effects will be analyzed in the next section.

### *C. Effects of AQM in Video Quality*

In this section, we analyze the impact of various AQM strategies deployed at the RLC buffer with different configuration parameters and we measure the QoE of a video streaming session in terms of VMAF and RD. Additionally, we also explore its impact on different QUIC implementations with multiple ABR schemes and CC algorithms.

Fig. 7 presents the average VMAF and RD analysis across five clients for scenario A (refer Table V), evaluating various AQM strategies for multiple server-side QUIC implementations—-MVFST, NGTCP2, QUINN, and S2NQUIC—-using BBR as the CC algorithm. On the client-side, QUIC-GO is deployed, leveraging Pensieve$^{+}$ as the ABR scheme.

For MVFST, ARED80, RED90, and L4S10-15 achieve comparable VMAF scores, indicating that these AQMs effectively regulate queue occupancy, maintaining stable throughput while minimizing packet loss. ARED80 dynamically adjusts *maxp* to maintain queue occupancy within the optimal range, RED90 relies on probabilistic early dropping at a relatively high threshold. Unlike its conventional ECN-based operation, L4S10-15, when used in non-ECN mode, actively drops packets at lower queue delays, ensuring tighter CC and reducing latency, which results in the highest VMAF among these AQMs. CODEL10 lags behind slightly in VMAF, as its aggressive dropping at lower queue delays sometimes disrupts steady throughput. DROPTAIL performs the worst, with the lowest VMAF, falling behind by $1\times$ the Just Noticeable Difference (JDN)[1] threshold. This degradation stems from its lack of any queue management strategy, leading to bursty packet losses when the buffer overflows, severely impacting video quality. Despite variations in VMAF, all AQM strategies exhibit minimal rebuffering.

For NGTCP2, we observe significantly lower VMAF scores ($\sim 2\times$ the Just Noticeable Difference (JND) threshold) compared to MVFST and with more rebuffering ($\sim 5\times$ more), regardless of the AQM strategy used. The primary reason for this disparity lies in the fundamental differences in how these implementations handle concurrent QUIC connections. MVFST leverages the *folly::AsyncTransport* library, which efficiently manages multiple QUIC clients asynchronously, while NGTCP2 relies on multi-threading to achieve concurrency. While the performance of both implementations may be comparable in non-congested scenarios, however, in our high-congestion environment with frequent queuing events in the RLC buffer, multi-threading introduces inefficiencies. Specifically, thread contention and synchronization overhead in NGTCP2 can lead to increased processing delays and suboptimal CC decisions. In contrast, MVFST's event-driven, asynchronous approach ensures smoother handling of concurrent connections, reducing packet processing delays and maintaining higher throughput. This enables MVFST to deliver higher-quality video representations with less rebuffering, resulting in superior VMAF scores compared to NGTCP2. For NGTCP2, ARED80 achieves the highest VMAF, surpassing other AQM strategies by $1\times$ the JND threshold. This can be attributed to ARED80's balanced congestion management, which minimizes unnecessary packet loss while maintaining higher throughput. Unlike more aggressive strategies like RED90 and L4S10-15, which react more quickly to congestion by marking or dropping packets, ARED80 allows moderate queue buildup, ensuring a steadier packet flow that benefits adaptive bitrate streaming. This contrasts with MVFST, where ARED80, RED90, and L4S10-15 exhibited comparable performance. By allowing slightly more queue buildup while preventing excessive losses, ARED80 mitigates the inefficiencies of NGTCP2's multi-threading model, enabling a higher throughput and ultimately yielding the highest VMAF.

For QUINN, we observe slightly higher VMAF scores compared to MVFST, with comparable rebuffering events and thus, QUINN exceeds NGTCP2 in VMAF by approximately $\sim 2\times$ the JND threshold while achieving $\sim 5\times$ less rebuffering. This improvement can also be attributed to QUINN's use of Rust's *async-tokio* library, which provides efficient asynchronous connection handling. By leveraging an event-driven architecture, QUINN minimizes blocking operations and thread contention, resulting in smoother congestion management and fewer disruptions in playback quality. While MVFST also employs an asynchronous approach, QUINN's Rust-based implementation offers enhanced memory safety and lower synchronization overhead compared to MVFST's C++ implementation. This results in more efficient connection handling and reduced contention under high congestion, giving QUINN a slight edge in VMAF performance in highly congested environments. With QUINN, ARED80 achieves the highest VMAF, outperforming the more aggressive RED90. The even more aggressive CODEL10 and L4S10-15 lag behind by $1\times$ the JND threshold compared to ARED80, as their increased packet dropping under congestion leads to slight quality degradation. DROPTAIL performs similarly to CODEL10 and L4S10-15 but with a slightly lower VMAF, further highlighting the trade-off between queue management aggressiveness and video quality.

[1] A difference of 6 units in VMAF

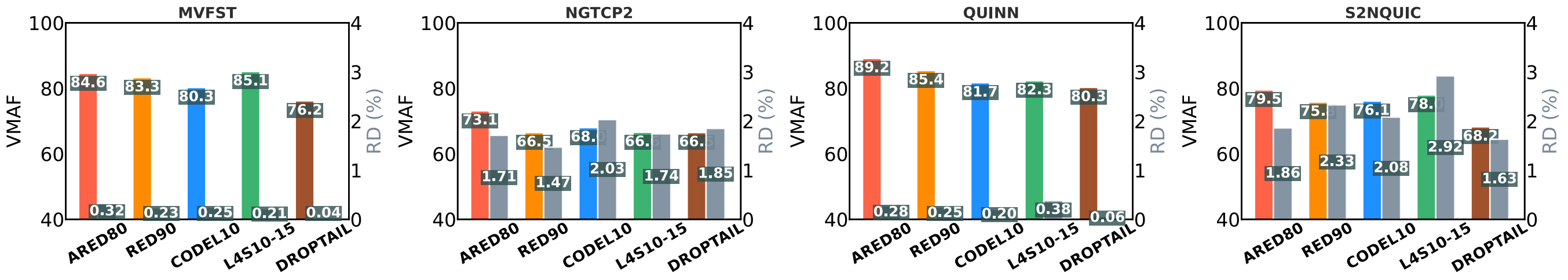

Fig. 7: QUIC BBR NON-ECN Pensieve.

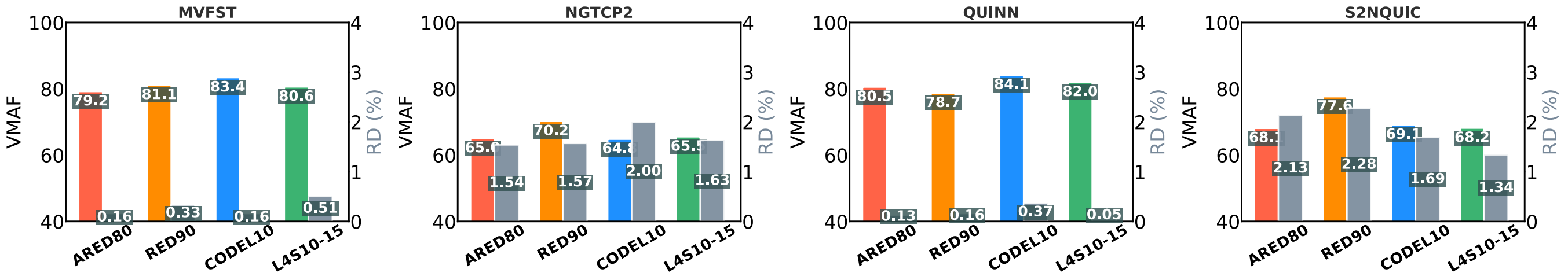

Fig. 8: QUIC BBR ECN Pensieve.

For S2N-QUIC, which also utilizes asynchronous connection management through Rust's *async-tokio* library—similar to QUINN—the performance is notably lower, with VMAF scores lagging by $1\times$ the JND threshold on average across all AQM strategies, and rebuffering events are about $10\times$ more frequent. Despite the efficient connection handling, the key factor driving this performance gap is the CC algorithm: while QUINN, MVFST, and NGTCP2 leverage the more advanced BBRv2, S2N-QUIC is limited to the older BBRv1. BBRv2 offers better responsiveness and adaptability to fluctuating network conditions through its dynamic bandwidth adaptation and delay-sensitive features, which allows for improved throughput and video quality. In contrast, BBRv1's less refined congestion management leads to greater packet loss, more rebuffering, and a slight degradation in VMAF scores. Despite this, S2N-QUIC outperforms NGTCP2 in terms of VMAF, exceeding by $1\times$ the JND threshold. This can be attributed to the efficient asynchronous connection management provided by S2N-QUIC, which contrasts with the multi-threaded approach used by NGTCP2, leading to less overhead and more efficient handling of concurrent connections. However, S2N-QUIC does experience more rebuffering compared to NGTCP2, which can be traced back to the lower delay-sensitivity of BBRv1 in S2N-QUIC. With S2N-QUIC, ARED80 delivers the highest VMAF, while RED90, CODEL10, and L4S10-15 perform comparably, slightly trailing behind ARED80. The use of BBRv1 in S2N-QUIC exacerbates the negative impact of the absence of queue management in DROPTAIL, leading to excessive queuing and congestion. As a result, DROPTAIL lags behind by $\sim 2\times$ the JND threshold compared to ARED80.

▷ **Key Takeaways** MVFST and QUINN (async-driven) outperform NGTCP2 (multi-threaded) and S2NQUIC (BBRv1-limited) in VMAF due to efficient congestion handling, with ARED80 balancing throughput and latency best. Aggressive AQMs (CODEL, L4S) risk quality via excessive drops, while DROPTAIL's lack of queue control causes severe VMAF loss. BBRv2 in QUIC implementations (vs. BBRv1 in S2NQUIC) reduces rebuffering, highlighting CC algorithm impact on QoE.

### D. *Effects of ECN marking in Video Quality*

Next, building upon the same experimental setup as in Section IV-C, we enable ECN marking (scenario C in Table V) across multiple network components, including the AQM strategies at the RLC buffer, the server-side QUIC implementations, and the router (or tc Netem) the results for which are illustrated in Fig. 8. Note that we observed similar results for ECN marking with scenarios C and D (refer Table V) with multiple ABR schemes and thus, for brevity, we only present the results for scenario C with Pensieve$^{+}$ as the ABR scheme. Notably, for DROPTAIL, since it lacks any form of queue management, ECN marking is not applicable. Instead, packets are either admitted or dropped outright when the buffer reaches capacity, leading to bursty losses rather than early congestion signaling. For MVFST, enabling ECN marking instead of packet dropping (non-ECN mode) results in a $\sim 1\times$ decrease in VMAF and a 50% reduction in RD. This behavior is expected, as ECN marking sets the CE bit in the IP header, signaling congestion before actual packet loss occurs. Consequently, MVFST reacts by reducing its congestion window earlier, leading to slightly lower throughput. While this helps in minimizing rebuffering, the trade-off is a marginal reduction in video quality in terms of VMAF due to the overall decrease in available bandwidth. We observe a similar trend with RED90, where VMAF slightly decreases with ECN marking, while rebuffering remains comparable to the non-ECN case. Although ECN marking helps mitigate severe congestion by proactively signaling congestion events, it does not always guarantee less rebuffering events. RED90 employs a static early dropping threshold with a higher probability of marking packets earlier, leading to more frequent congestion window reductions in ECN mode. This results in a slightly more pronounced decrease in throughput and marginally longer rebuffering durations due to the limited buffer capacity in LLL mode (6s). The results for L4S10-15 mirror those of RED90 due to the same primary factors. Interestingly, for CODEL10, the results in ECN mode are reversed compared to other AQM strategies. With ECN enabled, CODEL10 achieves slightly higher VMAF while also reducing rebuffering duration. This improvement can be attributed to CODEL10's intrinsic mechanism of dropping packets based on estimated standing queue delays. In non-ECN mode, CODEL10 aggressively drops packets once queue delays exceed a predefined threshold, sometimes disrupting steady throughput and leading to minor instability. However, in ECN mode, instead of outright dropping packets, CODEL10 marks them with ECN-CE, allowing the CC algorithm to respond more gradually rather than through abrupt packet losses. Additionally, the asynchronous nature of MVFST's connection management is particularly beneficial for CODEL10, which relies on precise queue delay estimation to trigger congestion responses. Since ECN enables early signaling without immediate packet drops, MVFST's low-latency, event-driven design allows it to adjust its congestion window more responsively, ensuring smoother throughput and reducing rebuffering events.

For NGTCP2, we observe a similar trend to MVFST when ECN marking is enabled, but with two key distinctions. First, RED90 shows a slight increase in VMAF, whereas CODEL10 experiences a minor decline. These differences stem from NGTCP2's multi-threaded architecture, which influences how CC reacts to ECN signals. RED90 benefits from NGTCP2's multi-threading because its probabilistic early dropping approach with ECN spreads congestion events more evenly across multiple threads, preventing sudden throughput collapses and ensuring smoother congestion adaptation. In contrast, CODEL10 relies on precise queue delay estimation, which is more effective when connections are managed asynchronously rather than across multiple competing threads. The higher synchronization overhead in NGTCP2's multi-threaded design results in delayed congestion responses, leading to suboptimal throughput adjustments and slightly lower VMAF for CODEL10 in ECN mode.

For QUINN, the results with ECN marking enabled closely resemble those of MVFST. This similarity can also be attributed to QUINN's asynchronous connection management, which is similar to MVFST.

For S2NQUIC, the results mirrored those observed for NGTCP2, with some notable trends in the presence of ECN marking. When ECN marking is enabled, ARED80 and L4S10-15 exhibit a VMAF decrease of nearly $2\times$ the JND threshold compared to non-ECN mode. This is due to the way ECN marking triggers congestion window reductions in S2NQUIC, which may slightly limit throughput and, in turn, affect video quality. However, RED90 in ECN mode yields a slight improvement in VMAF, while CODEL10 lags behind by about $1\times$ the JND threshold. This discrepancy can be attributed to the BBRv1 CC algorithm used in S2NQUIC, which is less delay-sensitive than BBRv2 used in other implementations like QUINN, MVFST, and NGTCP2. While BBRv1's lack of fine-tuned congestion management leads to higher queuing delays, this effect is not as pronounced with RED90 using ECN marking. RED90's probabilistic early packet drop mechanism is designed to manage congestion by marking packets, which combined with the larger queue length as compared to ARED80, allows it to signal congestion early without causing excessive packet loss, which is crucial for BBRv1, whose slower congestion response could otherwise exacerbate queuing delays.

$\triangleright$ **Key Takeaways** Enabling ECN marking reduces packet loss through early congestion signaling but trades marginally lower throughput (e.g., $1\times$ JND drop in VMAF for MVFST) for reduced rebuffering duration (50% lower RD). CODEL10 improves VMAF with ECN due to delay-sensitive marking aligning with async QUIC implementations (e.g., MVFST/QUINN), while static-threshold AQMs (RED90, L4S) exhibit mixed results, and DROPTAIL remains ineffective. Async-driven QUIC implementations (MVFST/QUINN) adapt better to ECN via responsive CC, whereas multi-threaded NGTCP2 and S2NQUIC (using BBRv1) suffer delays from synchronization overhead, with S2NQUIC showing $2\times$ JND VMAF degradation for ARED80/L4S due to BBRv1's sluggish response to ECN signals. BBRv1's limitations (e.g., RED90's slight VMAF gain but higher RD) underscore the critical role of CC algorithms in balancing ECN's throughput-latency trade-offs.

### *E. Effects of CC in Video Quality*

In this section, we analyze whether the impact of AQM strategies deployed at the RLC buffer layer on the QoE of video streaming follows the same trend despite the CC algorithm being used at the QUIC server. Following the same setup as in Section IV-C, we change the CC algorithm at QUIC server-side from BBR to CUBIC (scenario B in Table V), the results for which are highlighted in Fig. 9.

For MVFST, we observe significant drops in VMAF—by approximately 3× the JND threshold—along with substantially increased rebuffering duration, regardless of the AQM strategy used. This can be attributed to MVFST's reliance on CUBIC with HyStart, which struggles in our highly congested scenario. The frequent packet loss events cause CUBIC's congestion window to shrink drastically, leading to lower throughput and increased buffering. Additionally, as a loss-based CC algorithm, CUBIC does not adapt well to high RTT, further exacerbating performance degradation. As a result, all AQMs—ARED80, RED90, CODEL10, L4S10-15, and even DROPTAIL—experience lower VMAF and increased rebuffering. Notably, L4S10-15 exhibits minimal rebuffering since its aggressive packet-dropping at lower queue delays enforces tighter CC, reducing rebuffering more effectively than other AQMs.

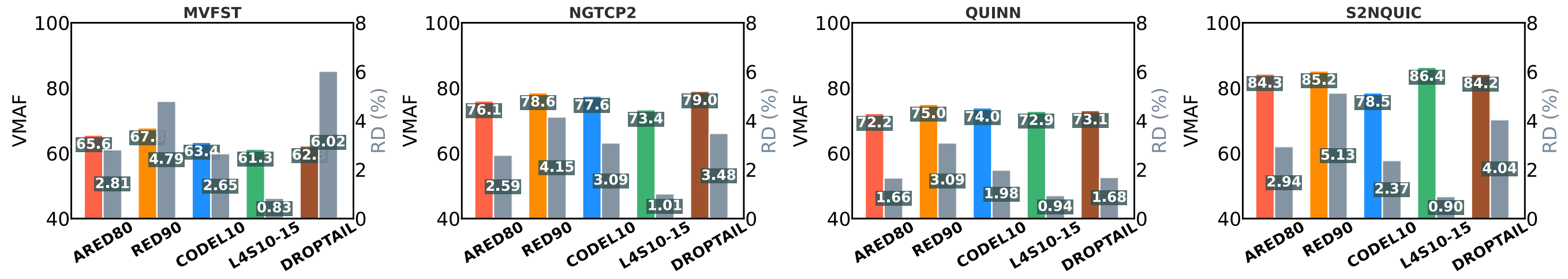

Fig. 9: QUIC CUBIC NON-ECN Pensieve.

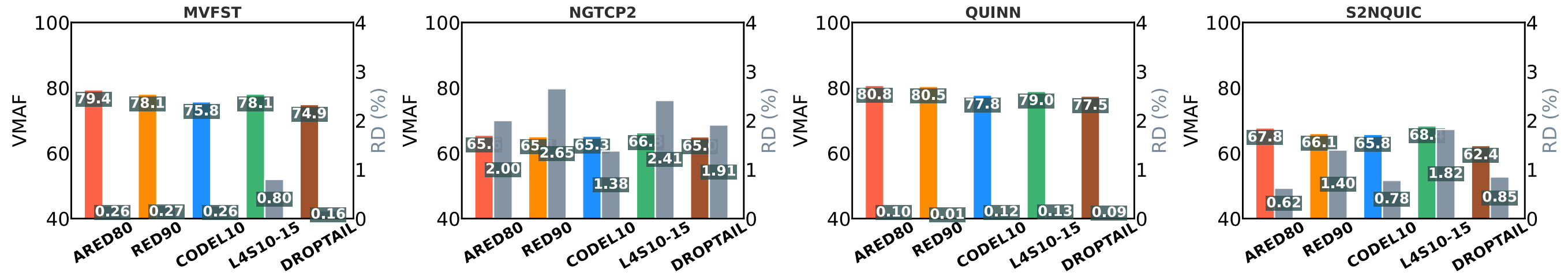

Fig. 10: QUIC BBR NON-ECN Conventional.

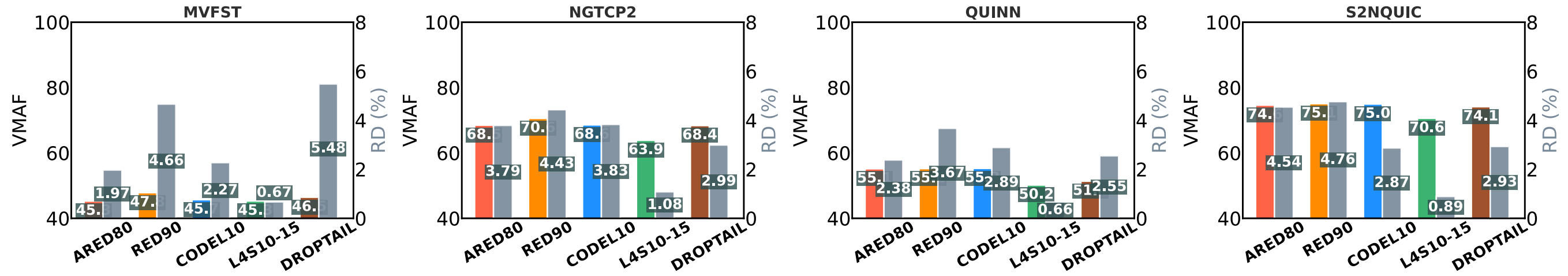

Fig. 11: QUIC CUBIC NON-ECN Conventional.

For NGTCP2, switching the CC algorithm from BBR to CUBIC produces the opposite effect compared to MVFST. While CUBIC in MVFST led to severe drops in VMAF and increased rebuffering, in NGTCP2, most AQMs see a substantial improvement in VMAF—by approximately 2× the JND threshold—albeit with slightly higher rebuffering in some cases. This divergence is primarily due to NGTCP2's use of CUBIC with HyStart++ (an upgrade to HyStart), which optimizes slow-start performance by utilizing RTT thresholds and mitigates excessive congestion window reductions in high-loss environments. ARED80 provides a slight improvement in VMAF compared to BBR as the CC algorithm. RED90, CODEL10, and DROPTAIL all demonstrate significantly higher VMAF compared to CUBIC in MVFST. While these AQMs introduce some additional rebuffering due to their more aggressive dropping behavior, they prevent prolonged congestion-induced queuing, which, when combined with CUBIC's more stable congestion window evolution in NGTCP2 due to Hystart++, results in overall increased throughput. L4S10-15 stands out as the only AQM that not only improves VMAF but also reduces rebuffering compared to BBR. Unlike other AQMs, which introduce packet loss events that can temporarily stall congestion window growth, L4S10-15's proactive early dropping maintains low queuing delay, ensuring that CC decisions occur at an optimal rate. This allows CUBIC in NGTCP2 to maintain a steadier congestion window without excessive oscillations, leading to both improved VMAF and reduced rebuffering.

For QUINN, switching the CC algorithm from BBR to CUBIC results in a performance trend similar to MVFST. Both QUIC implementations employ asynchronous connection management for handling multiple concurrent streams, which influences how congestion events propagate and how AQMs interact with CC. Unlike MVFST, which uses CUBIC with HyStart, QUINN does not incorporate HyStart and thus, QUINN's CC does not overreact to loss as aggressively as MVFST's CUBIC with HyStart, the overall QoE degradation remains slightly less pronounced. While this reduces overreaction to loss, the absence of HyStart may become more impactful as network congestion increases or as the number of hops between the server and client grows, potentially worsening performance over longer paths.

For S2NQUIC, the shift in CC algorithm from BBR to CUBIC follows the same pattern as NGTCP2, largely due to both using CUBIC with HyStart++. This helps maintain a steadier congestion window, preventing drastic throughput drops and resulting in an increased server-side throughput compared to MVFST and QUINN.

▷ **Key Takeaways** Switching from BBR to CUBIC as the CC algorithm significantly impacts QoE, with performance trends diverging across QUIC implementations. MVFST suffers severe VMAF degradation (3× JND drop) and increased rebuffering due to CUBIC's loss-based inefficiency in high congestion, while NGTCP2 and S2NQUIC (both using CUBIC with HyStart++) improve VMAF (up to 2× JND gain) via stabilized congestion windows, despite slight RD increases. QUINN (lacking HyStart) shows less severe VMAF drops than MVFST but underperforms due to weaker loss adaptation. AQM effectiveness varies: L4S10-15 in NGTCP2 reduces RD via proactive drops, while MVFST/QUINN's async models struggle with CUBIC's RTT sensitivity. The results highlight CC-AQM interdependencies—CUBIC's performance hinges on HyStart++ optimizations and AQM aggressiveness (e.g., L4S10-15 balances throughput-latency trade-offs better than static-threshold AQMs like RED90).

### *F. Effects of ABR in Video Quality*

In this section, we analyze whether the impact of AQM strategies deployed at the RLC buffer layer on the QoE of video streaming follows the same trend despite the ABR scheme being used at the QUIC client. Following the same setup as in Section IV-C, we change the ABR scheme at QUIC client-side from Pensieve$^{+}$ to CON$^{+}$, the results for which are depicted in Fig. 10.

For MVFST, the impact of AQM strategies observed with Pensieve$^{+}$ as the ABR scheme remains consistent for CON$^{+}$, with VMAF lagging by approximately 1× the JND threshold. This performance gap stems from the fundamental differences in how the two ABR schemes make decisions. CON$^{+}$ relies solely on the last two throughput estimations, making it more reactive and susceptible to short-term fluctuations, which can lead to suboptimal bitrate switches. In contrast, Pensieve$^{+}$, as a learning-based ABR scheme, processes multiple QoE-relevant metrics and considers the last eight observations as state inputs, allowing it to make more stable and adaptive bitrate decisions.

A similar trend is observed for NGTCP2 as well. When using CON$^{+}$ as the ABR, it results in reduced VMAF and increased rebuffering due to suboptimal bitrate guidance provided by the ABR. The results for QUINN and S2NQUIC show similar impacts from different AQM strategies, regardless of the ABR scheme used, with slightly lower VMAF. This reduction is solely due to the ABR decisions and does not alter the effect that each AQM has on the overall quality of the video streaming session. Notably, the reduced rebuffering for S2NQUIC when switching ABRs from Pensieve$^{+}$ to CON$^{+}$ is due to the client requesting fewer high-quality segments, a result of suboptimal bitrate decisions by CON$^{+}$. This, in turn, alleviates the negative impact of S2NQUIC using the older BBRv1 version.

Even while using the same setup as in Section IV-E, changing the ABR scheme at QUIC client-side from Pensieve$^{+}$ to CON$^{+}$, the results, as depicted in Fig. 11 follow the same pattern as changing the ABR scheme with with BBR as the CC algorithm.

▷ **Key Takeaways** Switching from Pensieve$^{+}$ to CON$^{+}$ reduces QoE across QUIC implementations (e.g., 1× JND VMAF drop for MVFST/NGTCP2) due to CON$^{+}$'s reliance on short-term throughput fluctuations, causing suboptimal bitrate decisions. AQM trends remain consistent, but CON$^{+}$'s lower-quality segment requests inadvertently reduce rebuffering for S2NQUIC (BBRv1-limited), masking congestion issues. QoE degradation persists with both BBR and CUBIC CC, emphasizing ABR's critical role in balancing AQM/CC dynamics—learning-based schemes (Pensieve$^{+}$) outperform rule-based (CON$^{+}$) by leveraging multi-metric, adaptive bitrate selection.

## V. Discussions

This study analyzed the impact of various AQM strategies at the RLC buffer on video streaming QoE, considering different QUIC implementations, ABR schemes, CC algorithms, and ECN marking. Our findings indicate that MVFST consistently achieves the highest VMAF, benefiting from its asynchronous connection handling and efficient congestion management. QUINN performs comparably, leveraging Rust's async-tokio library, while NGTCP2 exhibits lower QoE due to inefficiencies in its multi-threading model under congestion. S2NQUIC, despite using an asynchronous approach, is constrained by BBRv1, resulting in increased rebuffering. Among AQM strategies, ARED80 generally yields the best performance by balancing CC and queue occupancy, while aggressive dropping in CODEL10 and L4S10-15 leads to minor QoE degradation. DROPTAIL consistently underperforms due to bursty losses.

ECN marking plays a crucial role in mitigating severe congestion by proactively signaling network congestion before packet loss occurs, enabling CC algorithms to react earlier. However, this early signaling often results in a trade-off between throughput and video quality, as seen in the slight reduction in VMAF across various AQM strategies. While approaches like RED90 and CODEL10 benefit from smoother congestion adaptation, others such as ARED80 and L4S10-15 experience excessive congestion window reductions, leading to noticeable video quality degradation. The impact of ECN marking also varies across QUIC implementations, with asynchronous architectures like MVFST and QUINN leveraging early signaling for better congestion management, while multi-threaded designs like NGTCP2 face synchronization challenges that can result in suboptimal throughput adjustments. Ultimately, the effectiveness of ECN marking depends on both the AQM strategy and the CC algorithm in use, making it essential to carefully tune its deployment for optimal video streaming performance. It is noteworthy that in QUIC implementations, the processing of ECN marking is only treated as a normal congestion event for all packets marked with ECN. A specially designed ECN-aware CC implementation that intelligently adjusts the congestion window based on the proportion of packets marked and unmarked with ECN, while dynamically tuning the pacing rate, could hold significant promise for further advancements in ECN marking for QUIC. By leveraging finer-grained congestion signals, such an approach could enable smoother throughput adaptations, reducing unnecessary drops while maintaining low rebuffering.

The choice of CC algorithm, particularly BBR versus CUBIC, significantly affects the interaction with AQM strategies and consequently influences video streaming performance. BBR, being a delay-based CC algorithm, aims to maintain a steady bandwidth utilization by reacting to round-trip time (RTT) and congestion signals, which helps in stabilizing throughput under normal conditions. However, when paired with aggressive AQMs like RED90 or L4S10-15, BBR may not perform optimally. These AQMs tend to introduce packet drops or ECN markings during periods of congestion, potentially causing BBR's congestion window to decrease too quickly, resulting in throughput instability and rebuffering. In contrast, CUBIC, a loss-based CC algorithm, responds to packet loss by adjusting its congestion window size. While CUBIC typically shows worse performance under congested conditions, especially in highly lossy scenarios (as seen in MVFST), it behaves differently when paired with AQMs. For instance, CUBIC with HyStart++ (used in NGTCP2 and S2NQUIC) is more adaptive and optimized for high-loss environments. In this setup, AQMs like RED90, CODEL10, and L4S10-15 can provide better VMAF by ensuring that congestion-induced packet loss is minimized, and throughput is kept more stable. This results in smoother video playback and fewer rebuffering events compared to BBR. Our results underscore that the choice of a CC algorithm plays a key role in the impact of queuing management in the RLC buffer on the overall QoE of QUIC video streaming, and no single AQM is CC-agnostic. This highlights the significant potential for new AQM-aware CC designs that can dynamically optimize performance across different CC strategies.

Our results for different ABR schemes at the player side emphasize that the performance differences observed on the overall QoE of video streaming are more heavily dependent on the choice of ABR itself. Therefore, more intelligent cross-layer designs that consider metrics from the transport layer, application layer ABR metrics, and RLC link layer AQM metrics could open the door for more efficient and optimized video streaming experiences.

Our results underline the critical role of AQM strategies, QUIC implementations, ECN marking, CC algorithms, and ABR schemes in improving the viewer QoE, which is essential for operators to meet their Quality of Service (QoS) and Service Level Agreement (SLA) targets. Some possible solutions to further enhance QoE could involve dynamic AQM strategies combined with adaptive configuration of router parameters, allowing real-time adjustments based on traffic conditions. Additionally, cross-layer designs that integrate application, transport, and network layers could enable more responsive and efficient network management. Advanced techniques like machine learning-driven CC and real-time feedback loops between application and transport layers could further optimize resource utilization and QoE. While these solutions look promising, they require further investigation to assess their practicality, scalability, and real-world deployment in diverse network environments, marking a key direction for future research work.

Overall, the findings from our performance evaluation demonstrate that the effectiveness of AQM strategies and QUIC configurations is highly sensitive to the interaction between CC algorithms, RLC buffer dynamics, and ABR responsiveness under realistic 5G conditions. While certain combinations—such as asynchronous QUIC stacks with BBRv2 and moderately aggressive AQM (e.g., ARED)—consistently yield better QoE, these outcomes are not universally prescriptive. The optimal setup depends on the deployment context, including user mobility, radio conditions, and application requirements. Therefore, this work should be viewed as a generalizable framework that guides the design and evaluation of streaming strategies rather than a fixed recipe. Network operators and system designers are encouraged to adapt the parameters and components of this framework to align with their specific operational environments and performance objectives.

## VI. Conclusions and Future Work

Our paper explores the intricate interplay between AQM strategies, CC algorithms, ABR schemes, ECN marking, and QUIC implementations in optimizing video streaming QoE. We introduce a dual-phase evaluation framework: offline testing in 5G-LENA with various AQMs and an online Vegvisir extension for real-time streaming, incorporating diverse QUIC implementations, CC algorithms, and ABR schemes. This comprehensive approach provides a robust foundation for enhancing QoE in next-generation networks. Our findings underscore that no single AQM strategy is universally optimal. The selection of ABR schemes, CC algorithms, and QUIC implementations plays a critical role in performance outcomes. Additionally, ECN marking, when combined with intelligent CC, shows promise in alleviating congestion and improving video quality. Our study highlights the necessity of cross-layer optimization, integrating application-layer QoE insights with transport and data-link layer performance. This holistic approach enables adaptive strategies that dynamically respond to network conditions, reducing latency, enhancing stability, and improving user experience. Moreover, it equips network operators with actionable intelligence for efficient traffic management and resource allocation, ultimately improving network efficiency, reducing deployment costs and postponing capacity expansions.

Looking ahead, we plan to expand our investigation to include additional CC algorithms and ABR schemes while exploring advanced QUIC features such as browser rendering optimizations and adaptive image loading to further elevate streaming performance. Additionally, we aim to broaden our evaluation of AQM strategies and leverage machine learning techniques to enable more dynamic, intelligent network management—paving the way for enhanced scalability and efficiency across diverse network conditions.

## Acknowledgments

This work has been supported by the National Agency for Research and Development (ANID) / DOCTORADO

BECAS CHILE / 2021 - 21211859, MITACS Globalink Research Award IT38731, ANID Basal Project AFB240002, the Natural Sciences & Engineering Research Council (NSERC)-Discovery Grant RGPIN-2023-04744, NSERC Alliance (ALLRP 586434-23)-Mitacs Accelerate (IT36792) grant, and FRQNT Grant 343849.

## References

[1] Sandvine, "The Global Internet Phenomena Report, White Paper," *Technical Report, Sandvine*, January 2024.

[2] A. Bentaleb, B. Taani, A. C. Begen, C. Timmerer, and R. Zimmermann, "A survey on bitrate adaptation schemes for streaming media over HTTP," *IEEE Communications Surveys & Tutorials*, vol. 21, no. 1, pp. 562–585, 2018.

[3] "Transmission Control Protocol," RFC 793, Sep. 1981. [Online]. Available: https://www.rfc-editor.org/info/rfc793

[4] A. Langley, A. Riddoch, A. Wilk, A. Vicente, C. Krasic, D. Zhang, F. Yang, F. Kouranov, I. Swett, J. Iyengar *et al.*, "The QUIC transport protocol: Design and internet-scale deployment," in *ACM SIGCOMM*, 2017.

[5] S. Cook, B. Mathieu, P. Truong, and I. Hamchaoui, "Quic: Better for what and for whom?" in *2017 IEEE International Conference on Communications (ICC)*. IEEE, 2017, pp. 1–6.

[6] Y. Yu, M. Xu, and Y. Yang, "When quic meets tcp: An experimental study," in *2017 IEEE 36th International Performance Computing and Communications Conference (IPCCC)*. IEEE, 2017, pp. 1–8.

[7] M. Zhang, M. Polese, M. Mezzavilla, J. Zhu, S. Rangan, S. Panwar, , and M. Zorzi, "Will tcp work in mmwave 5g cellular networks?" *IEEE Communications Magazine*, vol. 57, pp. 65–71, 2019.

[8] R. Poorzare and A. C. Auge, "Fb-tcp: A 5g mmwave friendly tcp for urban deployments," *IEEE Access*, vol. 9, pp. 82 812–82 832, 2021.

[9] J. I. Sandoval and S. Céspedes, "Performance Evaluation of Congestion Control Over B5G/6G Fluctuating Scenarios," in *Proceedings of the Int'l ACM Symposium on DIVANET*, ser. DIVANet '23. New York, NY, USA: ACM, 2023, p. 85–92.

[10] D. Brunello, I. J. S, M. Ozger, and C. Cavdar, "Low Latency Low Loss Scalable Throughput in 5G Networks," in *2021 IEEE 93rd Vehicular Technology Conference (VTC2021-Spring)*, vol. 2021-April. IEEE, 4 2021, pp. 1–7. [Online]. Available: https://ieeexplore.ieee.org/document/9448764/

[11] A. Paul, H. Kawakami, A. Tachibana, and T. Hasegawa, "Effect of AQM-Based RLC Buffer Management on the eNB Scheduling Algorithm in LTE Network," *Technologies*, vol. 5, p. 59, 9 2017.

[12] M. Irazabal, E. Lopez-Aguilera, I. Demirkol, R. Schmidt, and N. Nikaein, "Preventing RLC Buffer Sojourn Delays in 5G," *IEEE Access*, vol. 9, pp. 39 466–39 488, 2021.

[13] J. I. Sandoval, S. Céspedes, A. González, D. Torreblanca, and I. Bugueño-Córdova, "A Deep Dive into Congestion Control and Buffer Management for Fluctuation-Prone 5G-A/6G Links," in *7th Conference on Cloud and Internet of Things. In press.* IEEE CIoT '24, 2024.

[14] K. D. Schepper and B. Briscoe, "The Explicit Congestion Notification (ECN) Protocol for Low Latency, Low Loss, and Scalable Throughput (L4S)," RFC 9331, Jan. 2023. [Online]. Available: https://www.rfc-editor.org/info/rfc9331

[15] K. Nichols, V. Jacobson, A. McGregor, and J. Iyengar, "Controlled Delay Active Queue Management," RFC 8289, Jan. 2018.

[16] E. Kinnear, J. Iyengar, and M. Thomson, "Quic loss detection and congestion control," RFC 9332, May 2023, dOI: 10.17487/RFC9332. [Online]. Available: https://datatracker.ietf.org/doc/html/rfc9332

[17] Q. W. Group. (2024) QUIC implementations. Accessed: 2024-07-21. [Online]. Available: https://github.com/quicwg/base-drafts/wiki/Implementations

[18] N. Cardwell, Y. Cheng, C. S. Gunn, S. H. Yeganeh, and V. Jacobson, "Tcp bbr v2 alpha/preview release," 2019.

[19] S. Ha, I. Rhee, and L. Xu, "Cubic: a new tcp-friendly high-speed tcp variant," *ACM SIGOPS operating systems review*, vol. 42, no. 5, pp. 64–74, 2008.

[20] H. Mao, R. Netravali, and M. Alizadeh, "Neural adaptive video streaming with pensieve," in *SIGCOMM*, 2017, pp. 197–210.

[21] TSGR, "TR 138 901 - V16.1.0 - 5G; Study on channel model for frequencies from 0.5 to 100 GHz (3GPP TR 38.901 version 16.1.0 Release 16)," 2020. [Online]. Available: https://www.etsi.org/deliver/etsi_tr/138900_138999/138901/16.01.00_60/tr_138901v160100p.pdf

[22] J. Sandoval and S. Cespedes, "Performance Evaluation of IEEE 802.11ax for Residential Networks," in *2021 IEEE Latin-American Conference on Communications (LATINCOM)*, 2021, pp. 1–7.

[23] D.-M. Chiu and R. Jain, "Analysis of the increase and decrease algorithms for congestion avoidance in computer networks," *Computer Networks and ISDN Systems*, vol. 17, pp. 1–14, 6 1989. [Online]. Available: https://linkinghub.elsevier.com/retrieve/pii/0169755289900196

[24] N. Cardwell, Y. Cheng, C. S. Gunn, S. H. Yeganeh, and V. Jacobson, "BBR: Congestion-based congestion control," *Communications of the ACM*, vol. 60, pp. 58–66, 2 2017.

[25] S. Ha, I. Rhee, and L. Xu, "CUBIC: A New TCP-Friendly High-Speed TCP Variant," *SIGOPS Oper. Syst. Rev.*, vol. 42, no. 5, p. 64–74, jul 2008. [Online]. Available: https://doi.org/10.1145/1400097.1400105

[26] S. Floyd and V. Jacobson, "Random early detection gateways for congestion avoidance," 1993.

[27] W.-C. Feng and D. D. Kandlur, "A Self-Configuring RED Gateway," 1999.

[28] S. Floyd, R. Gummadi, and S. Shenker, "Adaptive RED: An Algorithm for Increasing the Robustness of RED's Active Queue Management," 2001.

[29] K. Nichols and V. Jacobson, "Controlling queue delay," *Communications of the ACM*, vol. 55, pp. 42–50, 7 2012.

[30] E. Dumazet, "The Flow Queue CoDel Packet Scheduler and Active Queue Management Algorithm," 2018.

[31] K. Ramakrishnan, "RFC 3168 - The Addition of Explicit Congestion Notification (ECN) to IP," 2001.

[32] D. A. Alwahab and S. Laki, "ECN-Marking with CoDel and its Compatibility with Different TCP Congestion Control Algorithms," in *3rd International Conference on Advanced Science and Engineering, ICOASE 2020*. IEEE, 12 2020, pp. 74–79.

[33] W. Pan, H. Tan, X. Li, J. Xu, and X. Li, "Improvement of BBRv2 Congestion Control Algorithm Based on Flow-aware ECN," *Security and Communication Networks*, vol. 2022, 2022.

[34] I. Johansson and Z. Sarker, "Self-Clocked Rate Adaptation for Multimedia," RFC 8298, Dec. 2017. [Online]. Available: https://www.rfc-editor.org/info/rfc8298

[35] J. Herbots, M. Vandersanden, P. Quax, and W. Lamotte, "Vegvisir: A testing framework for http/3 media streaming," in *Proceedings of the 14th Conference on ACM Multimedia Systems*, 2023, pp. 403–409.

[36] K. Koutlia, B. Bojovic, Z. Ali, and S. Lagen, "Calibration of the 5G-LENA System Level Simulator in 3GPP reference scenarios," 2022.

[37] H. Yin, P. Liu, K. Liu, L. Cao, L. Zhang, Y. Gao, and X. Hei, "Ns3-ai: Fostering artificial intelligence algorithms for networking research," in *Proceedings of the 2020 Workshop on Ns-3*, ser. WNS3 2020. New York, NY, USA: ACM, 2020, p. 57–64. [Online]. Available: https://doi.org/10.1145/3389400.3389404

[38] L. Guo and I. Matta, "The war between mice and elephants," in *Proceedings Ninth International Conference on Network Protocols. ICNP 2001*, 2001, pp. 180–188.

[39] G. T. R. WG1, "Final Report of 3GPP TSG RAN WG1 #104-e v1.0.0," 4 2021. [Online]. Available: https://www.3gpp.org/ftp/tsg_ran/WG1_RL1/TSGR1_104b-e/

[40] B. Taraghi, H. Amirpour, and C. Timmerer, "Multi-codec ultra high definition 8k mpeg-dash dataset," in *Proceedings of the 13th ACM Multimedia Systems Conference*, 2022, pp. 216–220.

[41] T. Wiegand, G. J. Sullivan, G. Bjontegaard, and A. Luthra, "Overview of the h. 264/avc video coding standard," *IEEE Transactions on circuits and systems for video technology*, vol. 13, no. 7, pp. 560–576, 2003.

[42] R. Rassool, "Vmaf reproducibility: Validating a perceptual practical video quality metric," in *IEEE BMSB*. IEEE, 2017, pp. 1–2.

[43] M. Seemann, "Quic interop runner," https://interop.seemann.io/, 2025, automated interoperability testing of QUIC implementations.

[44] Y. Cheng, N. Cardwell, I. Järvinen, J. Mårtensson, I. Swett, and C. Wendt, "Hystart++: Modified slow start for tcp," IETF, Internet-Draft draft-ietf-tcpm-hystartplusplus-14, 2024, work in Progress. [Online]. Available: https://datatracker.ietf.org/doc/draft-ietf-tcpm-hystartplusplus/14/

[45] C. Huitema, "Multipath support in client: Can't change paths in video example," https://github.com/private-octopus/picoquic/issues/1758, May 2024, gitHub issue #1758, private-octopus/picoquic.

[46] i. SoonyangZhang, "It seems these functions in pacing.rs are not called," https://github.com/Tencent/tquic/issues/272, 2024.

[47] J. Herbots, A. Verstraete, M. Wijnants, P. Quax, and W. Lamotte, "Cross that boundary: Investigating the feasibility of cross-layer information sharing for enhancing abr decision logic over QUIC," in *NOSSDAV*, 2023, pp. 50–57.

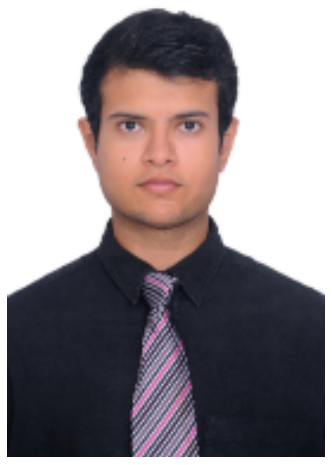

**Jashanjot Singh Sidhu** received his B.E. degree in Computer Engineering from Thapar Institute of Engineering and Technology, India. He worked as a software engineer at Samsung and later at VMware, gaining valuable industry experience in large-scale systems and networking. Since 2023, he has been pursuing a Ph.D. in Computer Science at Concordia University, Canada. His research interests include computer networking protocols, QUIC, multimedia streaming, applied AI for video streaming applications, and volumetric media delivery.

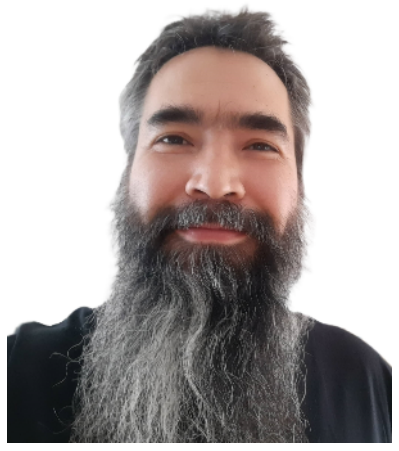

**Jorge Ignacio Sandoval** received a B.E. degree in Electrical Engineering and an M.Eng degree in Communication Networks from the University of Chile in 1997 and 2011, respectively, where he is currently pursuing a Ph.D. degree in Electrical Engineering. Since 2000, he has been an Instructor Professor in the Department of Electrical Engineering at the University of Chile. He has worked for more than 25 years in the industry, including 15 years at Chile's largest mobile operators, Telefónica and Entel, leading network planning and value-added services teams. His research interests include mobile networks, protocol designs, CC, and quality of experience.

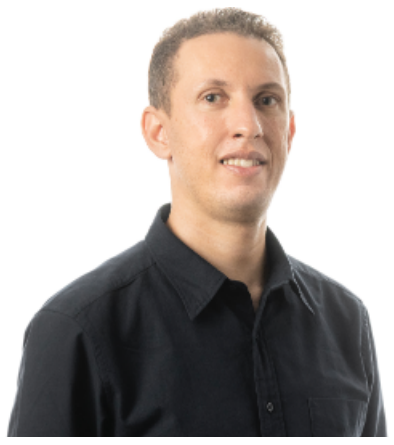

**Abdelhak Bentaleb** received his Ph.D. in computer science from the National University of Singapore (NUS), Singapore, in 2019. He continued as a research fellow at the same department until 2022. He is currently a a gina cody research and innovation fellow and assistant professor at the Department of Computer Science and Software Engineering, Concordia University, Canada. He is the founder and director of IN2GM Lab, Canada, a co-founder of Atlastream Inc., Singapore. He received many prestigious awards like SIGMM Award for Outstanding PhD Thesis Award, DASH-IF Best PhD Dissertation Award and Dean's Graduate Research Excellence Award AY2018/2019. His research interests include applied AI in multimedia systems and communication, video streaming architectures, content delivery, distributed computing, computer networks and protocols, wireless communications, and mobile networks. Further information can be found at https://users.encs.concordia.ca/~abentale/.

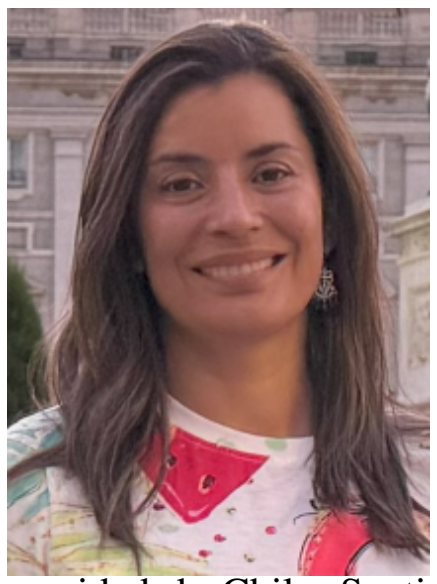

**Sandra Céspedes** received her B.Eng. (2003) and Specialization (2007) degrees in Telematics Engineering and Management of Information Systems from Universidad Icesi, Colombia, and her Ph.D. (2012) in Electrical and Computer Engineering from the University of Waterloo, Canada. Since 2022, she has been an Assistant Professor in the Department of Computer Science & Software Engineering at Concordia University, Montreal, Canada. Previously, from 2014 to 2021, she was an Associate Professor in the Department of Electrical Engineering at Universidad de Chile, Santiago, Chile. Dr. Céspedes is an Associate Researcher with the Advanced Center of Electrical and Electronic Engineering (AC3E) in Chile and an IEEE Senior Member. Her research focuses on mobile networking, protocol design for the Internet of Things, LEO satellite IoT networks, connected vehicles, and cyber-physical systems. She has served as an Associate Editor and TPC for several IEEE journals and conferences, including the IEEE Internet of Things Journal, IEEE Vehicular Technology Magazine, IEEE VNC, IEEE Globecom, and IEEE PMRIC, among others. Dr. Céspedes is an active participant and a co-author of several Internet standards at the Internet Engineering Task Force.